\documentclass[iop]{emulateapj}
\usepackage{amsmath}
\usepackage{threeparttable}

\slugcomment{Accepted for publication in ApJ}

\shorttitle{Simulating the Synthesis of Amino Acids in Meteorite Parent Bodies}
\shortauthors{Cobb et al.}

\usepackage{etex}
\usepackage{m-pictex}
\usepackage{m-ch-en}

\newcommand\textlcsc[1]{\textsc{\MakeTextLowercase{#1}}}

\begin{document}

\title{Nature's Starships II: Simulating the Synthesis of Amino Acids in Meteorite Parent Bodies}

\author{Alyssa K. Cobb\altaffilmark{1,2,3}, Ralph E. Pudritz\altaffilmark{1,2,4} and Ben K. D. Pearce\altaffilmark{1,2,5}}

\altaffiltext{1}{Origins Institute, McMaster University, ABB 241, 1280 Main St,  Hamilton, ON, L8S 4M1, Canada}
\altaffiltext{2}{Department of Physics and Astronomy, McMaster University, ABB 241, 1280 Main St, Hamilton, ON, L8S 4M1, Canada}
\altaffiltext{3}{E-mail: alyssacobb107@gmail.com}
\altaffiltext{4}{E-mail: pudritz@physics.mcmaster.ca}
\altaffiltext{5}{E-mail: ben.pearce@alumni.ubc.ca}

\begin{abstract}

Carbonaceous chondrite meteorites are known for having high water and organic material contents, including amino acids. Here we address the origin of amino acids in the warm interiors of their parent bodies (planetesimals) within a few million years of their
formation, and connect this with the astrochemistry of their natal protostellar disks. We compute both the total amino acid abundance pattern as well as the relative frequencies of amino acids within the CM2 (e.g. Murchison) and CR2 chondrite subclasses based on Strecker reactions within these bodies. We match the relative frequencies to well within an order of magnitude among both CM2 and CR2 meteorites for parent body temperatures $<$ 200$^{\circ}$C. These temperatures agree with 3D models of young planetesimal interiors. We find theoretical abundances of approximately 7x$10^5$ parts-per-billion (ppb), which is in agreement with the average observed abundance in CR2 meteorites of 4$\pm$7x$10^5$, but an order of magnitude higher than the average observed abundance in CM2 meteorites of $2\pm2$x$10^4$. We find that the production of hydroxy acids could be favoured over the production of amino acids within certain meteorite parent bodies (e.g. CI1, CM2) but not others (e.g. CR2). This could be due to the relatively lower NH$_3$ abundances within CI1 and CM2 meteorite parent bodies, which leads to less amino acid synthesis. We also find that the water content in planetesimals is likely to be the main cause of variance between carbonaceous chondrites of the same subclass. We propose that amino acid abundances are primarily dependent on the ammonia and water content of planetesimals that are formed in chemically distinct regions within their natal protostellar disks.


\end{abstract}

\section{Introduction}

One of the central questions about the origin of life is whether the delivery of exogenous water and
organics to planets may have ultimately enabled it.  Carbonaceous meteorites certainly deliver a rich mixture of biomolecules to the Earth today.  
These organic materials, including amino acids, could have formed in a variety of
environments such as the interstellar medium, protostellar disks out of which planets and stars form, and within the planetesimals and asteroids that built the rocky planets.   
It is well known that these extraterrestrial sources could have contributed a very significant fraction of the organics present on the young Earth's surface at the 
dawn of life \citep{Chyba1992}.   

A direct connection that we have with the conditions that created such early amino acids comes from
 their measured abundances and relative abundance patterns within meteorites. A good example is the extensive chemical analysis that has been 
done of carbonaceous chondrite meteorites such as the famous Murchison meteorite \citep{Cronin1983}. The parent bodies of meteorites namely planetesimals which are typically up to 50 - 100 km in size, 
were formed during the creation of the solar system.    Their composition consists of rock and ice together with various simple 
molecules (eg. HCN and ammonia) and organics (eg. formaldehyde) that were inherited from the icy, dusty materials in disks and interstellar gas.
Such bodies also contained various radionuclides whose decay
heated their interiors up to $150^{\circ}$C for several million years.  Given time and meteorite impacts, their organic content has been continually introduced to the planet \citep{Chyba1992,Sephton2002,BottaBada,Glavin2011}. This abundance pattern may have provided important constraints on the nature of the genetic code that ultimately appeared \citep{Higgs2009}.    

The formation of complex molecules is still not well understood. Experimental work by \citet{Lerner1993} has shown that carbonaceous chondrites like Murchison (a so-called CM2-type)
may be explained by a Strecker-type synthesis wherein aldehyde molecules (such as formaldehyde) in the presence of water, ammonia and HCN give rise
to amino acids (such as glycine). \citet{Lerner1993} shows that Strecker synthesis retains the observed highly deuterated amino acids found in Murchison, whose source ultimately is the D-enriched precursors in the interstellar medium.  Moreover, with few exceptions, it is found that the major amino acid that is produced in any reaction is the one expected from the associated aldehyde. Not all classes of amino acids are necessarily formed in such reactions however, 
and there is some evidence that the so-called n-$\omega$ species may have been formed by catalytic processes associated with so-called Fischer-Tropsch-Type (FTT) reactions \citep{Burton2012}.  

The observation of planetesimals and complex organic molecules in protostellar disks is now on the cusp of a revolution in our understanding with the advent of 
the ALMA observatory. As one example, it will enable observations of planetesimal systems formed in disks around other stars as well as the capability to detect carbon-bearing
species in the later debris disks \citep{Dent2014}. ALMA has also been able to detect glycolaldehyde (the precursor to the amino acid serine) in abundances 10--15 times less than methyl formate and 2--3 times more than ethylene glycol in the warm gas surrounding the protostellar binary IRAS 16293-2422 \citep{Jorgensen}. Such aldehydes would be incorporated into planetesimals and comets where they could play a crucial role in amino acid synthesis, as we will show.

The overarching goal of this work is to thermodynamically model the synthesis of amino acids inside meteoritic parent bodies and to begin to link this to the astrochemistry of their natal disks. \citet{Cobb2014} (hereafter Paper I) took a closer look at overall amino acid abundances and the relative frequencies of amino acids within specific meteorite subclasses. This comprehensive view of amino acid abundance data provided us with the constraints that we now use in this computational study to reproduce these observed abundance patterns.

Early theoretical work on the synthesis of amino acids in hydrothermal environments was done by \citet{Schulte1995}. Using principles of thermodynamics and Gibbs free energies, they showed mathematically that the synthesis of amino acids via a Strecker synthesis-like pathway is energetically favorable in hydrothermal conditions.
\citet{Amend1998}  used thermodynamic considerations and Gibbs free energies to produce a theoretical yield of amino acids in hydrothermal ecosystems. They found that the synthesis of 20 proteinogenic amino acids is energetically favorable in hot ($100^{\circ}$C) hydrothermal environments, as opposed to their synthesis at surface seawater conditions.   \citet{Schulte2004} shifted focus from hydrothermal systems to meteoritic parent bodies. Using theoretical compositions of carbonaceous chondrite parent bodies, they calculated that various organic materials present in carbonaceous chondrites could have been synthesized during aqueous alteration of the parent body. 

In this work, we investigate the theoretical synthesis of proteinogenic $\alpha$-amino acids via Strecker-type pathways inside meteoritic parent bodies. These bodies are important for amino synthesis since their interiors are heated to temperatures exceeding $100^{\circ}$C for more than 1 Myr due to radioactive heating, as shown by full 3D hydrodynamics simulations of their interiors \citep{Travis2005}. Moreover, the time scale for these reactions which synthesize amino acids is no longer than a few thousand years \citep{Peltzer1984}. Therefore it is a good assumption that these processes will have achieved thermodynamic equilibrium, making the Gibbs free energy minimization a desirable approach. Our model inherently assumes this chemical equilibrium, and outputs a series of amino acid concentrations which minimize the Gibbs free energy. We model parent body interiors using temperatures and pressures given in \citet{Travis2005} as well as cometary abundance data of organics for the initial concentrations in our model. 

We compare our modeled amino acid abundances to observed total abundances in different subclasses of carbonaceous chondrite, measured in parts-per-billion (ppb). Additionally, in Paper I we calculated the average relative frequencies of various amino acids normalized to glycine, measured in moles of amino acid/moles of glycine. The amino acid abundances measured in hydrolyzed (instead of unhydrolyzed) hot water extracts were used for these calculations due to a larger quantity of available data. These average frequencies have since been refined and standard deviations have been added. We note that the variation in the measured relative amino acid frequencies is greater than any discrepancies introduced from using amino acid abundances in hydrolyzed meteorite extracts. We calculate the pattern of relative frequencies in theoretical yield, and match it to observation in two carbonaceous subclasses, CM2 and CR2.

In this work, we limit our investigation to proteinogenic $\alpha$-amino acid abundances. This is primarily because the thermochemical database with which we work currently only contains data for the 20 proteinogenic amino acids encoded by the universal genetic code---with the exception of one non-proteinogenic $\alpha$-amino acid. It is important to note that there are appreciable quantities of non-proteinogenic amino acids (such as amino-isobutyric acid, isovaline and $\alpha$-aminobutyric acid) and non-$\alpha$-amino acids (such as $\beta$-alanine) detected in carbonaceous chondrites. In Appendix I, we analyze one case: $\alpha$-aminobutyric acid. In addition, we currently only consider amino acids which are synthesized via Strecker-type reactions involving aldehydes. Amino acids such as amino-isobutyric acid and isovaline, while both highly abundant in carbonaceous chondrites, are synthesized via Strecker-type reactions with ketone precursors.

Our results give very good agreement with the observed data base of amino acid frequencies. In trying to understand the absolute abundances of amino acids in the various meteorite classes predicted in our models, we propose that the variations in the water and ammonia content of planetesimals formed at varying radii within natal protoplanetary disks would naturally account for these differences.  This is a new picture that is quite different than the current
"onion" structure of differentiated chemistry within a planetesimal.  

The layout of this paper is to first quickly review our observational results 
from Paper I, summarize the theoretical and computational methods that we used, and then to discuss our results and comparison to the 
data for both the abundances and relative frequencies of our models.   Finally, we discuss the results in a larger framework and introduce our new picture
in the last section of the paper.   

\section{Compilation of Meteoritic Amino Acid Data}

Paper I presented an extensive collation of available meteoritic data on amino acid abundances and relative frequencies in carbonaceous chondrites. Carbonaceous chondrites are the class of meteorite that contain the highest content of both water and organic materials, including amino acids. These abundances are seen to vary 
systematically with meteoritic type. In this section we give a quick summary of meteorite classification, followed by a summary of their amino acid abundances
(the reader is encouraged to read Paper I for the details).

\subsection{ Meteorite Classification - a Quick Guide}

Meteorites are assigned a number 1--6 indicating petrologic type, in addition to a subclass designation CI, CM, CR, CV, CO, CH, CB, and CK. Petrologic type carries with it information regarding degrees of internal chemical alteration undergone by a meteoritic parent body. Petrologic type 1 chondrites are the most aqueously altered. This means the liquid water  within the parent bodies has chemically altered their makeup. Petrologic type 2 meteorites have undergone significant degrees of aqueous alteration, though not to the extent of type 1 meteorites. Petrologic type 3 meteorites most closely resemble the solar nebulae in which they formed; they have undergone relatively little to no degrees of alteration. All carbonaceous chondrites subclasses from petrologic types 1--3 are known to contain high concentrations of water and organics. Petrologic types 4--6 indicate increasing degrees of thermal metamorphism. We restrict our modeling to meteorites of type 1--3, as samples in this regime have been aqueously altered and produced amino acids via Strecker synthesis.

\begin{figure*}
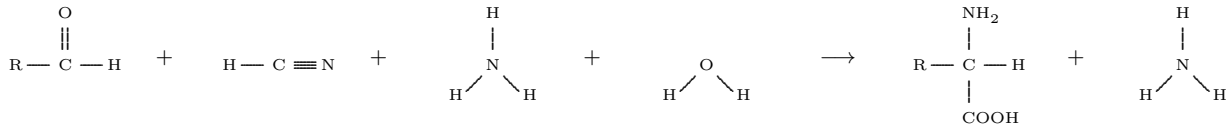

\hspace{-2mm}\startchemical\chemical[ONE,Z0157,SB15,DB7][C,H,R,O] \stopchemical
\hspace{-2mm}\chemical{PLUS} \hspace{-2mm} \startchemical \chemical[ONE,Z015,TB1,SB5][C,N,H] \stopchemical
\hspace{-2mm}\chemical{PLUS} \hspace{-2mm} \startchemical \chemical[ONE,Z0247,SB247][N,H,H,H]\stopchemical
\hspace{-2mm}\chemical{PLUS} \hspace{-2mm} \startchemical \chemical[ONE,Z024,SB24][O,H,H]\stopchemical 
\hspace{-0mm} \chemical{GIVES} 
\hspace{-0mm}\startchemical \chemical[ONE,Z01357,SB1357][C,H,COOH,R,NH_2] \stopchemical
\hspace{-2mm} \chemical{PLUS}\hspace{-2mm} \startchemical\chemical[ONE,Z0247,SB247][N,H,H,H]\stopchemical
\caption{General form of a Strecker-type synthesis reaction. An aldehyde (R-CHO) reacts with hydrogen cyanide (HCN) and ammonia (NH$_3$) in aqueous solution (H$_2$O) to produce an amino acid (H$_2$N-RCH-COOH) and a new ammonia molecule. R denotes the side chain, which varies for different aldehydes/amino acids. \label{Strecker}}
\end{figure*}

CI meteorites, all petrologic type 1, are among the most aqueously altered carbonaceous chondrites. They are generally associated with a higher range in temperatures, $50 ^{\circ}$C to $150 ^{\circ}$C, according to \citet{Krot2006}, \citet{Zolensky1993}, \citet{Huss2006}, and \citet{Grimm1989}, and they rank approximately fourth highest for average total amino acid abundances by carbonaceous chondrite subclass.

The CM and CR subclasses express the highest total amino acid concentrations, greater by several orders of magnitude than any other subclass. They also express the greatest variety of proteinogenic amino acids. These subclasses generally correspond to petrologic type 2, with several exceptions, and are associated with alteration temperatures in the broad range of $0 ^{\circ}$C to $240 ^{\circ}$C. CM-types are given a range of $0 ^{\circ}$C to roughly $50 ^{\circ}$C in some studies \citep{Zolensky1993,Huss2006,Krot2006,Grimm1989}, but one study assigns a much higher peak metamorphic temperature of $240 ^{\circ}$C \citep{Busemann}. Peak metamorphic temperatures of chondrites are typically estimated from phase changes or reactions that occur at specific temperatures \citep{Huss2006}. \citet{Busemann} on the other hand use the correlation between the D band width (obtained from Raman spectroscopy) and peak metamorphic temperature to base their estimates. CR meteorites, spanning petrologic types 1--3, also have a discrepancy in classification temperatures. \citet{Busemann} give a peak metamorphic temperature of $240 ^{\circ}$C, \citet{Huss2006} cite an upper temperature bound of 150--300$^{\circ}$C, and \citet{Zolensky1993} state a range of $50^{\circ}$C to $150^{\circ}$C. \citet{Morlok2013} provide a broader range in classification temperatures, $25^{\circ}$C and $300^{\circ}$C, associated with ``all stages of aqueous alteration.'' 

CV, CO, CH and CB meteorites are all petrologic type 3. These meteorites are associated with peak metamorphic temperatures ranging from 200--600$^{\circ}$C \citep{Busemann,Huss2006}, though \citet{Zolensky1993} has CV3 peak temperatures at a much lower 50$^{\circ}$C. These chondrites are thought to be the most pristine samples of material we have from the time of the solar system formation. They are relatively unaltered, either via aqueous alteration or thermal metamorphism---which occurs at higher temperatures of perhaps $\sim 500^{\circ}-950^{\circ}$C \citep{Huss2006} with no water present. With the exception of CH meteorites which are the third most abundant subclass in total amino acids, we observe low levels of amino acids in these meteorites. 

Lastly, the CK meteorites span petrologic types 3--6 and are associated with temperatures from 250--600$^{\circ}$C \citep{Burton2015}. A recent study has found one CK3 meteorite to be comparable in amino acid abundances to the CI1 types \citep{Burton2015}. The CK4--6 meteorites in the same study were found to have very low abundances of amino acids.

\subsection{Amino Acid Abundances}

In Figure 8 of Paper I, we plot average amino acid abundances in parts-per-billion (ppb) across different carbonaceous chondrite subclasses. The histogram style abundance plot shows average amino acid abundance for various proteinogenic amino acids for each of the six subclasses of carbonaceous chondrite we consider. This allows us to identify graphically in which subclasses there is a greater concentration of amino acids. The CO3 and CM1/CR1 sections have average abundances on the order of $10^1$ ppb. The average abundance of each amino acid for CV3 and CI1 subclasses is on the order of $10^2$ ppb. CM2-types contain $10^3$ ppb per amino acid on average. The CR2-types show the greatest variation in abundance, each average ranging from $10^3$--$10^5$ ppb. The CH3, CB3 and CK3 types were not included in Figure 8 of paper I, but have since been found to contain $10^2$--$10^3$ ppb, $10^1$--$10^2$ ppb and $10^2$ ppb per amino acid on average, respectively \citep{Burton2013,Burton2015}. From these two studies and Figure 8 of Paper I, we identified the petrologic type 2 meteorites, both CM2 and CR2, as containing greater total abundances of amino acids than either the petrologic type 1 or type 3 meteorites. This is in agreement with observations and conclusions by, for example, \citet{Martins2007}, \citet{Glavin2011}, and \citet{Burton2012}. 

\subsection{Amino Acid Frequencies}

In addition to modeling the total amino acid yield, we investigate the relative frequencies of amino acids.  Figures 9 and 10 in Paper I show the relative frequencies of 13 amino acids relative to glycine in the CM2 and CR2 subclasses, respectively. As the CM2 and CR2 subclasses contain the greatest concentrations and variety of amino acids, we selected these groups as a starting place for thermodynamic modeling. We normalize to glycine as this amino acid is classically one of the most abundant found in meteorites, it is the easiest to synthesize and is most energetically favorable, and it is achiral. In both subclasses, alanine is the second most abundant amino acid and glutamic acid is the third. Among the CM2 chondrites, there are also significant quantities of serine, proline, lysine, threonine, valine, isoleucine, leucine, aspartic acid, tyrosine, and phenylalanine. The CR2 samples show great variety in frequency; valine, isoleucine, leucine, proline, and phenylalanine are fairly prevalent, and there are small quantities of aspartic acid, threonine, serine, and tyrosine.  We use the same theoretical model used for calculating total amino acid yield to measure relative frequencies. 

\section{Theoretical Background}
\subsection{Strecker Chemistry}

The general form of a Strecker-type reaction is shown by the mass-balance equation in Figure~\ref{Strecker}.
Following the experimental work discussed in the Introduction \citep{Lerner1993}, we simplify our equilibrium chemistry scheme to focus on 
the relation between an aldehyde and its main amino acid. 
A Strecker-type reaction is defined as the reaction between an aldehyde (R-CHO), hydrogen cyanide (HCN) and ammonia (NH$_3$) in aqueous solution (H$_2$O) which produces an amino acid (H$_2$N-RCH-COOH) and an ammonia molecule. There is an intermediate step not depicted in Figure~\ref{Strecker}; in the Strecker synthesis of glycine, for example, reactants initially combine to create an aminoacetonitrile (NH$_2$CH$_2$CN), which then undergoes a one-way process called hydrolysis, resulting in the creation of glycine (C$_2$H$_5$NO$_2$). This general form may be used to define the underlying Strecker pathway for each of the proteinogenic amino acids laid out in Paper I. In each case, the side chain (R group) of a reactant aldehyde will be identical to the R group in the corresponding amino acid. 

\begin{deluxetable}{p{35mm}p{35mm}}
\tablecaption{Proteinogenic amino acids and their requisite Strecker aldehyde pairings. Includes the name and molecular formula of each amino acid and aldehyde.}
\centering
 \tablehead{\colhead{Amino Acid}&\colhead{Aldehyde}}
 \startdata
  glycine \newline C$_2$H$_5$NO$_2$&formaldehyde \newline CH$_2$O \\[1mm]
  alanine \newline C$_3$H$_7$NO$_2$ & acetaldehyde \newline C$_2$H$_4$O \\[1mm]
  valine \newline C$_5$H$_{11}$NO$_2$ & isobutyraldehyde \newline C$_4$H$_8$O \\[1mm]
  leucine \newline C$_6$H$_{13}$NO$_2$ & 3-methyl butanal \newline C$_5$H$_{10}$O \\[1mm]
  isoleucine \newline C$_6$H$_{13}$NO$_2$ & 2-methyl butanal \newline C$_5$H$_{10}$O \\[1mm]
  serine \newline C$_3$H$_7$NO$_3$ & glycolaldehyde \newline C$_2$H$_4$O$_2$ \\[1mm]
  lysine \newline C$_6$H$_{14}$N$_2$O$_2$ & 5-aminopentanal \newline C$_5$H$_{11}$NO \\[1mm]
  threonine \newline C$_4$H$_9$NO$_3$ & 2-hydroxy-propanal \newline C$_3$H$_6$O$_2$ \\[1mm]
  tyrosine \newline C$_9$H$_{11}$NO$_3$ & p-hydroxyphenyl-acetaldehyde C$_8$H$_8$O$_2$\\[1mm]
  phenylalanine \newline C$_9$H$_{11}$NO$_2$ & phenyl-acetaldehyde \newline C$_8$H$_8$O \\[1mm]
  aspartic acid \newline C$_4$H$_7$NO$_4$ & 3-oxopropanoic acid \newline C$_3$H$_4$O$_3$ \\[1mm]
  glutamic acid \newline C$_5$H$_9$NO$_4$ & 4-oxobutanoic acid \newline C$_4$H$_6$O$_3$ \\[1mm]
  histidine \newline C$_6$H$_9$N$_3$O$_2$ & imidazole-4-acetaldehyde \newline C$_5$H$_6$N$_2$O\\[1mm]
  arginine \newline C$_6$H$_{14}$N$_4$O$_2$ & N-(4-oxobutyl)guanidine \newline C$_5$H$_{11}$N$_3$O  \\[1mm]
  tryptophan \newline C$_{11}$H$_{12}$N$_2$O$_2$ & indol-3-ylacetaldehyde \newline C$_{10}$H$_9$NO\\[1mm]
  asparagine \newline C$_4$H$_8$N$_2$O$_3$ & 3-oxopropanamide \newline C$_3$H$_5$NO$_2$\\[1mm]
  glutamine \newline C$_5$H$_{10}$N$_2$O$_3$ & 4-oxobutanimide \newline C$_4$H$_7$NO$_2$\\[-3mm]
 \enddata \label{aldehydes}
\end{deluxetable}

This precise treatment of Strecker-type synthesis necessitates the R-group equivalence of the reactant and the product. For each Strecker reaction producing a proteinogenic amino acid, we can balance the incoming atom count with the outgoing atom count, which allows for the definitive determination of the aldehyde required to create an amino acid. Table~\ref{aldehydes} shows this information for 17 of the 20 proteinogenic amino acids: glycine (gly), alanine (ala), aspartic acid (asp), glutamic acid (glu), isoleucine (ile), leucine (leu), lysine (lys), serine (ser), threonine (thr), valine (val), tyrosine (tyr), phenylalanine (phe), histidine (his), arginine (arg), tryptophan (trp), glutamine (gln), and asparagine (asn). We exclude proline (pro) due to its inconformity with the basic Strecker-type treatment, as well as cysteine (cys) and methionine (met), which each contain a sulfur atom. For each listed amino acid, Table~\ref{aldehydes} provides the amino acid name and molecular formula as well as the name 
and molecular formula of the requisite Strecker aldehyde.

The information contained within Table~\ref{aldehydes} may be used to construct the unique Strecker-type reactions resulting in the synthesis of each of 17 proteinogenic amino acids. Aldehyde and amino acid pairings may be input into the generic Strecker reaction shown in Figure~\ref{Strecker}. For example, the simplest Strecker pathway is for glycine, shown by Eq. \ref{glycine} below.
\begin{align}\label{glycine}
\hspace{1mm}\chemical{CH_2O}
\hspace{1mm}\chemical{PLUS}\hspace{1mm} &\chemical{HCN} \hspace{1mm}\chemical{PLUS} \hspace{1mm}\chemical{NH_3}\hspace{1mm}\chemical{PLUS}\hspace{1mm}\chemical{H_2O}\hspace{3mm}  
\chemical{GIVES} \nonumber \\
& \hspace{1mm}\chemical{H_2N-HCH-COOH}
\hspace{1mm}\chemical{PLUS}\hspace{1mm}\chemical{NH_3}
\end{align}
Here the reactant is formaldehyde (CH$_2$O) and the product amino acid is glycine (H$_2$N-HCH-COOH, also seen in its collapsed form C$_2$H$_5$NO$_2$ in Table~\ref{aldehydes}). The R group in both the aldehyde and the amino acid is a single H atom. Note the traditional molecular structures shown in Figure~\ref{Strecker} have been collapsed to their molecular formulae for ease of reading.

\subsection{Gibbs Free Energies}

Thermodynamically speaking, a chemical reaction is favorable if the Gibbs free energy of reaction is negative. Each Strecker-type reaction we consider has some Gibbs free energy of reaction associated with it ($\Delta G_r$). In addition to species-specific Gibbs energies of reaction, there is a Gibbs free energy of formation ($\Delta G_f$) associated with each species involved in a reaction. $\Delta G_r$ may be computed at various temperature and pressure combinations using the $\Delta G_f$ for each species. These values are related via the relationship 
\begin{equation}
\Delta G_r=\Sigma G^{products}_f - \Sigma G^{reactants}_f.
\end{equation}

The values of $\Delta G_f$ change with temperature and pressure for each species according to Eq. \ref{coeff}. 
\begin{equation} \label{coeff}
  G_f(T,P) = a + bT + cTln(T) + dT^2 + eT^3 + f/T + gP
 \end{equation}
which provides us with the Gibbs coefficients a through g for each chemical species. Each term in this expression is derivable from the general definition of Gibbs free energy. Here, temperature (T) must be in Kelvin and pressure (P) in bars, and Gibbs energies return in Joules/mole. Eq. \ref{coeff} is the requisite Gibbs energy function as required by our modeling software, discussed in Section \ref{chemapp}.

\section{Computational Methods}
 
Our general strategy follows the outline described by Gibbs energies above. Initially we compute a series of Gibbs coefficients a--g for each chemical species involved in the Strecker-type synthesis of amino acids. Following this, we perform a minimization of the Gibbs free energy for each individual reaction in the entire network of Strecker equations. These two tasks are discussed in Sections \ref{chnosz} and \ref{chemapp}, respectively.

\subsection{Computing Gibbs Formation Energies}\label{chnosz}
 
We use a chemical thermodynamics database called CHNOSZ (version 0.9-9 (2013-01-01), authored by Jeffrey M. Dick, http://www.chnosz.net/). CHNOSZ contains Gibbs energies of formation for a wide variety of amino acids and aldehydes, in addition to water, ammonia, and hydrogen cyanide, all requisite species involved in the Strecker-type synthesis of amino acids in meteoritic parent bodies.

For each of the chemical species we are interested in (reactants and products), CHNOSZ contains a set of $\Delta G_f$ which change with respect to both temperature and pressure, spanning 0 to 1000$^{\circ}$C and 1 to 5000 bar. We plot these values and perform a least squares regression. We model the resulting line of best fit as in Eq. \ref{coeff}.

\begin{figure}[ht!]
 \centering
 \includegraphics[width=80mm]{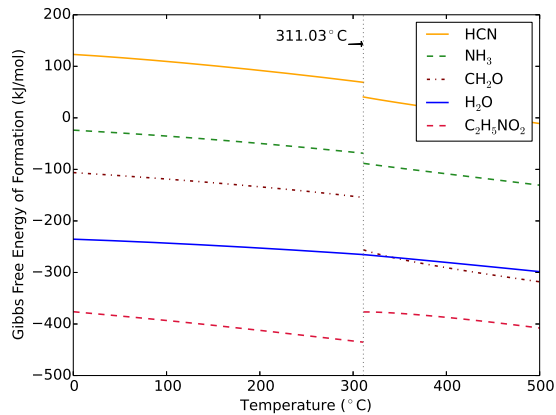}
 \caption{A plot of the reactant and product Gibbs energies of formation for the Strecker synthesis of glycine at 100 bar. The vertical grey dotted line represents the boiling point of water at 100 bar.}
 \label{phasechange}
 \end{figure}

Note that all reactant and product Gibbs energies in a single Strecker reaction simulation change abruptly after the boiling point of water for a given pressure. Figure~\ref{phasechange} illustrates this for the reactants and products of the Strecker synthesis of glycine at 100 bar. In this plot, hydrogen cyanide, formaldehyde, ammonia and glycine are discontinuous at 311.03$^{\circ}$C, and also the curve representing water---albeit difficult to see at this scale---slightly decreases in slope after this temperature. These curves represent a first-order phase transition; as their first derivatives are discontinuous. Glycine jumps to a higher Gibbs energy of formation after the liquid-to-gas phase transition of the mixture, corresponding to a decrease in energetic favourability. Conversely hydrogen cyanide, formaldehyde and ammonia become more favourable after the liquid-to-gas transition due to their discontinuous decrease in Gibbs energies.

\begin{figure}[ht!]
 \centering
 \includegraphics[width=80mm]{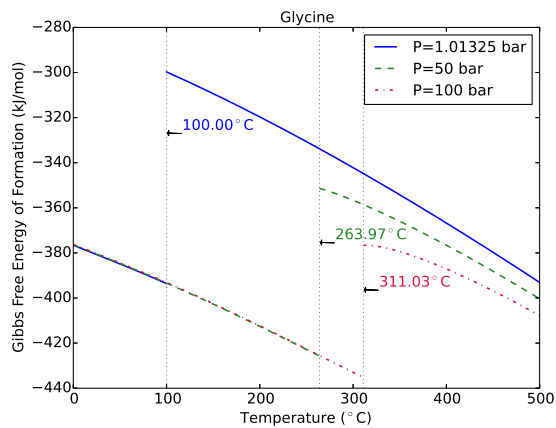}
 \caption{An example plot showing Gibbs energy dependence on pressure and temperature for the amino acid glycine. Temperature is allowed to vary from 0$^{\circ}$C to 500$^{\circ}$C. Three curves are shown, representing the Gibbs energies at pressures of 1.01325 bar (blue curve), 50 bar (green curve), and 100 bar (red curve). Note the discontinuities at the boiling point of water for each pressure.}
 \label{pressure}
 \end{figure}

Figure~\ref{pressure} shows the effect of varying pressure on the Gibbs energy of glycine. Note how each point of discontinuity is always at the boiling point of water for the respective curve's pressure. There is also an obvious lack of dependence on pressure before each curve reaches its liquid-to-gas phase transition; three pressure curves are shown (1.01325 bar, 50 bar, 100 bar), and they vary by approximately half a kilojoule from 0--100$^\circ$C. (Pressure hardly affects aqueous solution Gibbs free energies due to the fact that water is nearly an incompressible fluid.) Therefore, in our computations, we set the pressure to a static value of 100 bar in order to keep the theoretical mixture in the aqueous phase for a larger temperature range (as the aqueous phase is a requirement of Strecker synthesis). This pressure value is on the higher end of interiors of planetesimals, and is in good agreement with values cited by various authors, including \citet{Dodd1981}, \citet{Schulte1995}, \citet{Cohen2000}, and \citet{Warren2012}. In Appendix I, we show the Gibbs free energy dependencies on pressure and temperature for the other nine simulated amino acids (and $\alpha$-aminobutyric acid).
 
Our thermodynamic modeling software (see Section \ref{chemapp}), requires the set of coefficients a through f for each chemical species, as from Eq. \ref{coeff}. CHNOSZ contains the data necessary to calculate the coefficients for most of the species in our system, with the exception of glycolaldehyde. We have data for all other species used in our model: initial reactants (water, hydrogen cyanide, and ammonia), five aldehydes (formaldehyde, acetaldehyde, butyraldehyde, propionaldehyde, and pentanaldehyde), and the 17 amino acids listed in Table~\ref{aldehydes}. Glycolaldehyde was modeled as a mixture of acetaldehyde and acetic acid, as in \citet{Espinosa2005}. We obtain the associated Gibbs coefficients for glycolaldehyde using the technique in \citet{Emberson2010}.

It is important to note that the CHNOSZ database does not contain Gibbs formation energies for all the aldehydes listed in Table~\ref{aldehydes}. CHNOSZ only contains Gibbs data for 10 aldehydes, all differing by one carbon atom (the first, formaldehyde with one carbon atom, and the last, decanal with 10 carbon atoms). We have specific coefficients for formaldehyde, acetaldehyde, and glycolaldehyde, which are the Strecker precursors for glycine, alanine, and serine, respectively. 

All additional aldehydes in Table \ref{aldehydes} are modeled using the closest related aldehyde for which CHNOSZ contains available data. The three substitute aldehydes we use are butyraldehyde, propionaldehyde, and pentanaldehyde. These three, combined with formaldehyde, acetaldehyde, and glycolaldehyde, provide us with a range of aldehyde species which contain successive numbers of carbon atoms. The carbon count is the primary feature used to match Strecker aldehydes with available species substitutes. For example, the Strecker aldehyde necessary to synthesize valine is isobutyraldehyde, for which CHNOSZ does not contain data. We model the synthesis of valine using butyraldehyde rather than 
isobutyraldehyde, as it is a close substitute and contains an identical number of each atom. 

Aldehyde substitutions are used for val, ile, leu, lys, asp, glu, and thr, and are selected from the available aldehyde choices based on which contains the most similar molecular makeup. We require that the substitute aldehyde contain, at a minimum, an identical number of carbon atoms. We have made some quantitative estimates on how these aldehyde substitutions affect the results of our Strecker synthesis reactions in Appendix II. Synthesis simulations for gly, ala and ser, for which the specific requisite aldehyde Gibbs data is used, could be considered the most accurate. We use butyraldehyde for val and pentanaldehyde for both leu and ile. The molecular makeups of the substitutes match identically the actual Strecker aldehydes in these cases (C$_4$H$_8$O for butyraldehyde, C$_5$H$_{10}$O for pentanaldehyde). The substitutes aldehydes for lys, asp, glu, and thr contain the same number of carbon atoms as their actual aldehydes. We use pentanaldehyde for lys, propionaldehyde (C$_3$H$_6$O) for asp and thr, and butyraldehyde for glu. Our analysis in Appendix II shows that our results are reasonably accurate for all but two amino acids, glutamic acid and threonine, for which sensitivity to initial conditions could be important.

\subsection{Minimizing Gibbs Free Energy for the System: Thermodynamic Modeling}\label{chemapp}

We use a modeling software called ChemApp (distributed by GTT Technologies, http://www.gtt-technologies.de/newsletter) to perform a global minimization of Gibbs energy and calculate a theoretical yield of amino acids. ChemApp is a computational subroutine database of chemical thermodynamics which functions assuming a thermodynamic equilibrium of the system in question. For these systems, Gibbs free energy is minimized for a closed chemical system in equilibrium. The model simulates reactions in meteoritic parent bodies at temperatures and pressures that should theoretically exist in parent body interiors as a consequence of radionuclide decay. Thermal evolution models using radionuclide decay as a heat source have also been considered by \citet{Grimm1989}.

The functional structure of ChemApp is to read-in a data file containing the relevant molecular information for a given trial, including thermochemical data for all species (coefficients a--f calculated from CHNOSZ) and the temperature ranges of validity. The basic components of every file are the series of basic elements involved in our research: carbon, hydrogen, nitrogen, and oxygen. Each chemical species we consider is defined as the addition of its elemental components. We build a program which reads in the data file, and define a set of ChemApp subroutines at specific internal conditions including temperature, pressure, and chemical composition. In the case that reactants have far larger concentrations than the products, it is safe to assume weak coupling between the reactants and do single reaction calculations. For this reason our ChemApp system consists of several individual Strecker reactions. Each individual Strecker reaction is simulated by minimizing the total Gibbs free energy of the system using the thermochemical input from each individual reaction data file. Each
ChemApp simulation includes all species involved in the specific Strecker reaction, and is created using Figure~\ref{Strecker} and Table~\ref{aldehydes}. As already indicated, some simulations include aldehyde substitutions. These substitutions are necessary due to the unavailability of thermochemical data for some aldehyde species. ChemApp's Gibbs energy minimization is constrained by mass balance, and returns a set of molecular concentrations from the entire pool of reactants and products (one concentration value for each molecular species) which minimizes Gibbs free energy of the system.

It is important to note that ChemApp has no inherent method for differentiating between molecular structures. Because each input species is broken down to its elemental form, two ChemApp trials run with different molecular abundances, but identical elemental abundances, would return equivalent output concentrations. Additionally, our current treatment assumes an ideal solution for each reaction, and we only allow the chemistry to be in a single phase---aqueous or gaseous---at any given temperature and pressure.
 
Our current system restricts the possible outcomes of ChemApp to include only those species in which we are interested. Sources of initial reactants, e.g., hydrogen cyanide and ammonia, as well as alternate Strecker products, are not considered. This means, for example, we do not include the formation pathways which lead to the formation of aldehydes in ChemApp; we assume some initial aldehyde concentration. Also, although we do not explicitly consider the hydroxy acids which Strecker synthesis may also produce in these calculations, we do address their effects later in the paper (see Section~\ref{chemcomp}). In this way, we have a limited set of initial species and a limited set of possible products. ChemApp output reports the resulting abundances of amino acids and Strecker reactants which minimize Gibbs energy, without considering alternate products which may detract from amino acid abundances.

For amino acid synthesis calculations, ChemApp receives as input the data file containing the Gibbs coefficients a--f we calculated in Eq. \ref{coeff} for all species considered in a given run. Because the Gibbs free energy is independent of pressure until after the liquid-to-gas phase transition, and we are only interested in amino acid synthesis via aqueous phase reactions, we set the pressure to a value on the upper boundary of typical parent body interiors---keeping the simulation in the aqueous phase for a wide temperature range---and hold it constant.
Factors which may then be varied between ChemApp runs are temperature and the initial concentrations of reactants.
 
\subsection{Internal Conditions of Planetesimals} 

Our current ChemApp treatment includes a set pressure at 100 bar and a temperature range from $0^{\circ}$C to $500^{\circ}$C in increments of $25^{\circ}$C. \citet{McSween2002} and \citet{Travis2005} model thermal evolution and internal heating processes in meteoritic parent bodies of carbonaceous chondrites. The temperatures we use are in agreement with \citet{Travis2005} for the parent bodies of less altered carbonaceous chondrites. The higher temperatures follow the temperature classifications as laid out in numerous studies, including \citet{Sephton2002}, \citet{Weisberg2006}, and \citet{Taylor2011}, who discuss the temperature ranges associated with different subclasses of carbonaceous chondrite. Our temperature ranges are also in agreement with peak alteration temperatures among carbonaceous chondrite petrologic groups in \citet{Weiss2013}.

The internal conditions for our fiducial model are in agreement with cases 1--3 of the modeled carbonaceous chondrite parent bodies proposed by \citet{Travis2005}; that is, a spherical rocky body with a porosity of 20\%, a radius of 50 km and a rock density of 3000kg/m$^3$; initially with ice water of density 917kg/m$^3$ completely filling the pores of the body.

We are primarily interested in the effects of varying temperature, organic composition and water on amino acid synthesis. As discussed in Paper I, different subclasses of carbonaceous chondrite may be linked to different parent body formation conditions and therefore different amino acid abundance levels. We vary temperature in our ChemApp runs to recreate the temperature range across meteorite classifications.

Table \ref{initial} lists the values and ranges of initial concentrations of organics input into ChemApp. These values and ranges were selected as a starting place based on concentrations of organics found in comets. The molecules for which we have concentration data include ammonia, hydrogen cyanide, formaldehyde, acetaldehyde, and glycolaldehyde; all normalized to percent water. These concentrations are in agreement with cometary data in \citet{Bockelee2000}, \citet{Ehrenfreund2000}, \citet{Bockelee2004}, \citet{Crovisier2004}, and \citet{Mumma2011}, which include data from a variety of radio and infrared surveys. 

\begin{deluxetable}{p{35mm}p{35mm}}
\tablecaption{Initial concentrations of organics used in ChemApp model. Concentrations reported in percent normalized to water. Values are in agreement with \citet{Bockelee2000}, \citet{Ehrenfreund2000}, \citet{Bockelee2004}, \citet{Crovisier2004}, and \citet{Mumma2011}.\label{initial}}
\centering
 \tablehead{\colhead{Molecular Species}&\colhead{Concentration ([X]/[H$_2$O])}}
 \startdata
  Water & 100\\
  Ammonia & 0.7\\
  Hydrogen cyanide & 0.25\\
  Formaldehyde & 0.066\\
  Acetaldehyde & 0.02--0.08\\
  Glycolaldehyde & 0.0005--0.04\\
  Butyraldehyde & 0.0005--0.04\tablenotemark{*}\\
  Propionaldehyde & 0.0005--0.04\tablenotemark{*}\\
  Pentanaldehyde & 0.0005--0.04\tablenotemark{*}\\[-3mm]
 \enddata 
 \tablenotetext{*}{Estimated ranges.}
\end{deluxetable}

Cometary concentrations were chosen as a starting point for the reactants within our model planetesimal instead of carbonaceous chondrite concentrations for several reasons. Comets are thought to be the most unmodified bodies in the solar system \citep{Rauer}, and molecules in comets could have also been available to planetesimals at the time of the latter's formation \citep{Schulte2004,Alexander}. Carbonaceous chondrite matrices on the other hand are not thought to be pristine, being depleted in volatiles to varying degrees \citep{Bland2005}. Carbonaceous chondrites are also from parents bodies that have undergone high degrees of aqueous alteration \citep{Cobb2014}, and are for a number of reasons more susceptible to weathering than other meteorite types \citep{Bland2006}; making it difficult to obtain accurate parent body molecule fractions to water. Therefore it is safe to assume that cometary concentrations may be more likely to represent the molecular concentrations in a planetesimal during the formation of the solar system.

There are a few additional aldehydes included in our model for which we have no available cometary data. These are butyraldehyde, propionaldehyde, and pentanaldehyde. In these cases,
we match their initial concentrations with the ranges of initial concentrations for glycolaldehyde. We deem this to be an acceptable estimate because of the greater complexities of these aldehydes compared to glycolaldehyde.

Our first model treats a theoretical parent body as chemically homogeneous; the initial concentrations of reactants are uniform throughout the model parent body. 

\section{Results - Total Abundances}

\subsection{Dependence on Temperature}

\begin{figure*}
\centering
\includegraphics[width=\textwidth]{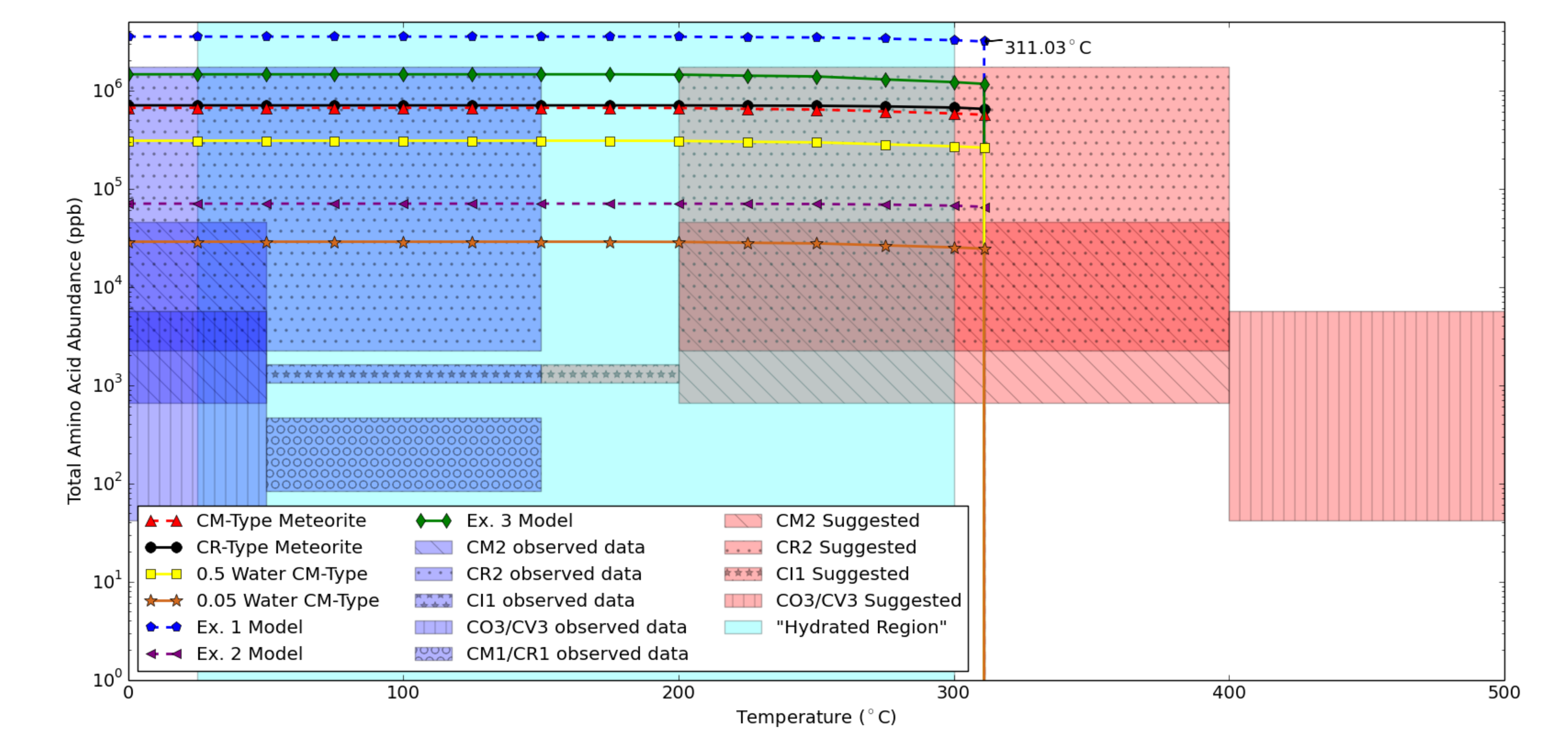}
\caption{Theoretical total yield of amino acids in ppb, for a variety of models. Each data point on a line represents a modeled total amino acid abundance at a given temperature, in intervals of $25^{\circ}$C. Amino acids included in each total concentration value are gly, ala, ser, asp, glu, val, thr, leu, ile, and lys. The shaded boxes indicate the observed range in amino acid abundance for each carbonaceous chondrite subclass (CI1, CM1/CR1, CM2, CR2, CV3/CO3). Dark blue shaded regions indicate subclass classification temperatures in agreement with \citet{Zolensky1993,Huss2006,Krot2006,Grimm1989}. The aqua shaded region indicates classification temperatures suggested in \citet{Morlok2013}. Red shaded regions correspond to temperature boundaries suggested by \citet{Sephton2002}.}
\label{comparison}
\end{figure*}

We use the physical model described above to compute the total theoretical yield of amino acids at temperature and pressure conditions analogous to those in the interior of a meteoritic parent body. The results of different models are shown in Figure \ref{comparison}. 

In Figure \ref{comparison}, data points from each model are shown by markers located at intervals of $25^{\circ}$C for the temperature range $0^{\circ}$C to $500^{\circ}$C and joined by a connecting line. Each data point represents the theoretical total amino acid yield. At each temperature point for each model, ChemApp returns some concentration value for individual amino acids; we plot the total amino acid content. The modeled total amino acid yield at each temperature is a sum of the following amino acid abundances: gly, ala, ser, asp, glu, val, thr, leu, ile, and lys. 

The different model curves include one for a parent body of CM-like composition, and one for a parent body of CR-like composition. In two other models, also with CM-like composition, we scale down the water---and therefore the corresponding organics---by one half and one twentieth respectively. Then there are another three models with CR-like composition which correspond to Ex. 1, Ex. 2, and Ex. 3, in which we alter the relative abundances of initial aldehydes from the CR column of Table \ref{CMRconcs}. Ex. 1 was run with aldehyde concentrations increased by an order of magnitude, Ex. 2 with aldehyde concentrations decreased by an order of magnitude, and Ex. 3 with a disrupted aldehyde pattern: formaldehyde, acetaldehyde, and butyraldehyde concentrations are equal to the values shown in the CR column of Table \ref{CMRconcs}, and glycolaldehyde, propionaldehyde, and pentanaldehyde concentrations have been increased by an order of magnitude.

\begin{deluxetable*}{cccccc}
\tablecolumns{6}
\tablecaption{Concentrations of aldehydes used in CM-type and CR-type relative frequency models which minimize $\chi^2$, and three experimental models of aldehyde concentrations (Ex. 1, 2 and 3).\label{CMRconcs}}
 \tablehead{\colhead{Molecular Species}&\colhead{CM ([X]/[H$_2$O])}&\colhead{CR ([X]/[H$_2$O])}&\colhead{Ex. 1 ([X]/[H$_2$O])}&\colhead{Ex. 2 ([X]/[H$_2$O])}&\colhead{Ex. 3 ([X]/[H$_2$O])}}
 \startdata
formaldehyde & 0.0660 & 0.0660 & 0.6600 & 0.0066 & 0.0660\\
acetaldehyde & 0.0350 & 0.0726 & 0.7260 & 0.0073 & 0.0726\\
butyraldehyde & 0.0073 & 0.0114 & 0.1140 & 0.0011 & 0.0114\\
glycolaldehyde & 0.0112 & 0.0066 & 0.0660 & 0.0007 & 0.0660\\
propionaldehyde & 0.0042 & 0.0011 & 0.0110 & 0.0001 & 0.0110\\
pentanaldehyde & 0.0066 & 0.0021 & 0.0210 & 0.0002 & 0.0210\\
 \enddata
\end{deluxetable*}

The observed amino acid abundances are shown as shaded regions in Figure~\ref{comparison}. The upper and lower boundaries of each shaded region enclose the range of observed abundances for a carbonaceous chondrite subclass (CI1, CM1/CR1, CR2, CM2, CV3/CO3). The left and right boundaries of each shaded region are the suggested temperature classifications for the corresponding subclass. Blue shaded regions cover temperature classifications in good agreement with the work done by \citet{Zolensky1993,Huss2006,Krot2006,Grimm1989}. The aqua shaded region indicates the broad classification boundaries cited by \citet{Morlok2013}, in which aqueous alteration may occur in hydrated regions between temperatures of $25^{\circ}$C and $300^{\circ}$C. Red shaded regions show subclass classification boundaries suggested by \citet{Sephton2002}. 

 \begin{figure}[ht!]
 \centering
 \includegraphics[width=80mm]{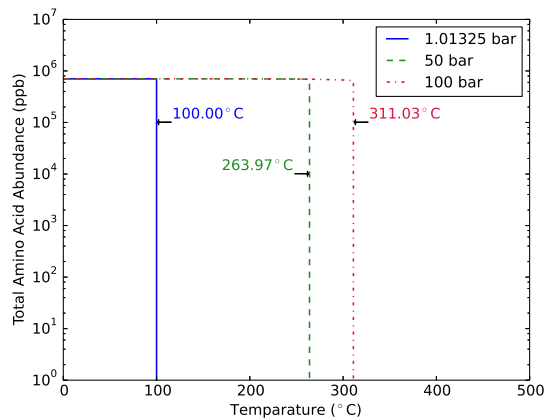}
 \caption{Theoretical total abundances of amino acids for a parent body of CR-like composition, simulated at different pressures. The blue, green and red curves represent the total abundances at 1.01325 bar, 50 bar and 100 bar respectively. The three labeled temperatures (100.00$^\circ$C, 263.97$^\circ$C and 311.03$^\circ$C) represent the boiling point of water at each of these three pressures.}
 \label{pressureProfile}
 \end{figure}

We observe good agreement between our CR2-type thermodynamic model and observed amino acid abundances for CR2 meteorites. In fact, all models except Ex. 1 lay within the range of observed amino acid abundances for CR2 meteorites. We observe an approximate order of magnitude difference between our CM2-type thermodynamic model and observed amino acid abundances for CM2 meteorites. The total amino acid abundances for the CR2- and CM2-type models are practically equivalent. This reveals a lack of sensitivity in the resultant total amino acid abundances to slight variations in the initial aldehyde concentrations. Theoretical abundances remain stable as temperature increases from 0$^\circ$C until about 200$^{\circ}$C---where the curves begin to decrease very slowly until they reach the drop-off point at the liquid-to-gas phase transition. 

Figure \ref{pressureProfile} illustrates how the pressure of the thermodynamic system plays a direct role in the temperature ranges at which aqueous alteration can occur. Aqueous alteration in petrologic types 1--3 occurs up to $300^{\circ}$C, as suggested by \citet{Morlok2013}. Since the liquid-to-gas phase change turns off amino acid synthesis---as made clear by the abrupt vertical drop-off of each curve in Figure \ref{pressureProfile} at their respective boiling points of water---the suggested limit of aqueous alteration by \citet{Morlok2013} may be a product of pressures not exceeding 100 bar within most meteorite parent bodies. On the other hand, \citet{Sephton2002} suggests that petrologic type 2 meteorites can synthesize amino acids up to $400^{\circ}$C. In order for our simulation to yield non-zero abundances of amino acids at this temperature, the initial pressure condition necessary would need to surpass the estimated pressure near the core of 10 Hygiea (the 4th largest asteroid in the Solar System) of 246 bar; as calculated by \citet{Bardon2013}.

\subsection{Dependence on Water Content}

A possible reason why our CM2-type thermodynamic model returned abundances an approximate order of magnitude greater than observed amino acid abundances in CM2 meteorites, is that different subclasses of meteorite could originate from different parent bodies, some of which tend to have less water. To exemplify these scenarios, the 0.5 and 0.05 water CM-type models in Figure~\ref{comparison} calculate total amino acid abundances with one half and one twentieth the water content in their planetesimals with respect to the other models. The 0.5 water CM-type model decreases in total amino acid abundance by approximately a factor of 3 compared to the CM-type model, but still does not reach the boundary of the observed CM2 meteoritic abundances. The 0.05 water CM-type model on the other hand decreases in total abundance by over an order of magnitude compared to the CM-type model and is well within the range of observed CM2 meteoritic abundances. 

ChemApp is only able to assume a uniform composition. Running separate ChemApp simulations with different levels of water and organics---pertaining to each meteorite subclass in our study---might help recover the overall observed amino acid abundance trends we see among the various carbonaceous chondrite subclasses.

\section{Results - Relative Frequencies}

\subsection{CM2 Meteorites}

\begin{figure*}
\centering
\includegraphics[width=\textwidth]{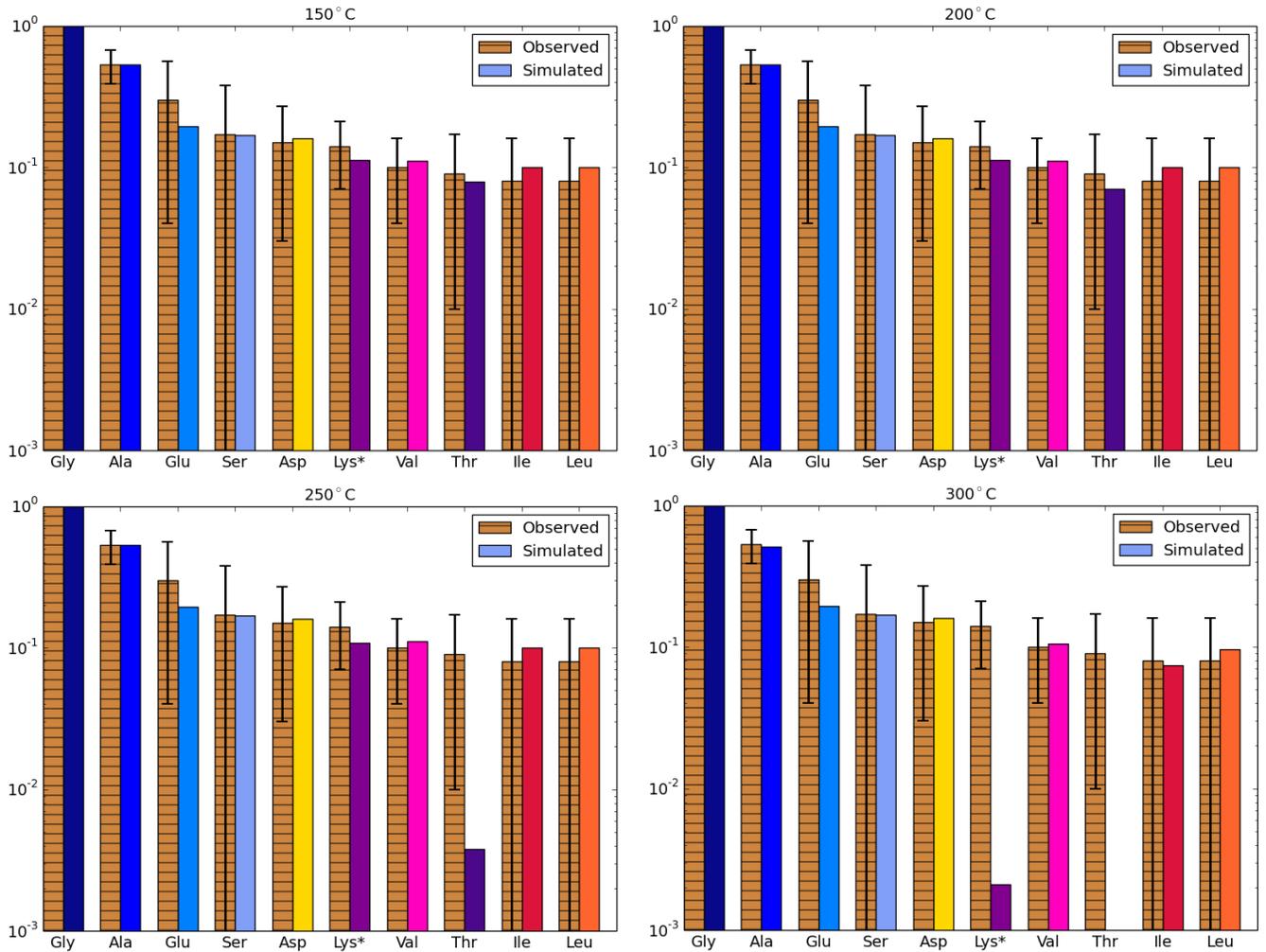}
\caption{Relative frequencies of amino acids in the CM2 subclass. Abundances of 10 amino acids are shown relative to glycine, reported in moles of amino acid/moles of glycine. Observed frequencies are shown in brown with black horizontal stripes, and black error bars representing the standard deviation. The modeled frequencies are placed beside the observed frequencies with a unique solid color representing each amino acid. The four panels represent the progression of relative amino acid abundances with temperature, starting at $150^{\circ}$C and increasing by $50^{\circ}$C until $300^{\circ}$C. *Limited meteoric data was available for averaging the observed frequency.}
\label{CMfreq}
\end{figure*}

Figure~\ref{CMfreq} shows the change in relative frequencies of 10 amino acids with temperature in the CM2 subclass. Relative frequencies are shown in moles of amino acid/moles glycine. The amino acids have been arranged in order of decreasing observed frequency. Each panel shows (in brown with black horizontal stripes) observed frequencies of amino acids relative to glycine among CM2 meteorites. Beside these observed frequencies stand (each in an individual color) the simulated relative frequencies of amino acids with respect to glycine. Each subsequent panel depicts the progression of simulated relative frequencies with temperature. We note that $150^{\circ}$C was selected as an arbitrary initial temperature, representative of petrologic type 2 meteorites. As discussed in the previous section, our model does not recreate the low abundance pattern at cooler temperatures, and therefore cannot differentiate between amino acid abundances from $0^{\circ}$C to $150^{\circ}$C.

The ranges of initial concentrations of aldehydes going into the model were taken directly from---or estimated using---observations of cometary abundances, as discussed above (see Table~\ref{initial}). In an effort to constrain formation conditions within a planetesimal, we performed a chi-squared minimization to optimize the concentrations of aldehydes down to single values. The concentrations producing the lowest chi-squared value (to four decimal places), used in the model producing Figure~\ref{CMfreq}, are shown in the CM column of Table~\ref{CMRconcs}.

\begin{figure*}
\centering
\includegraphics[width=\textwidth]{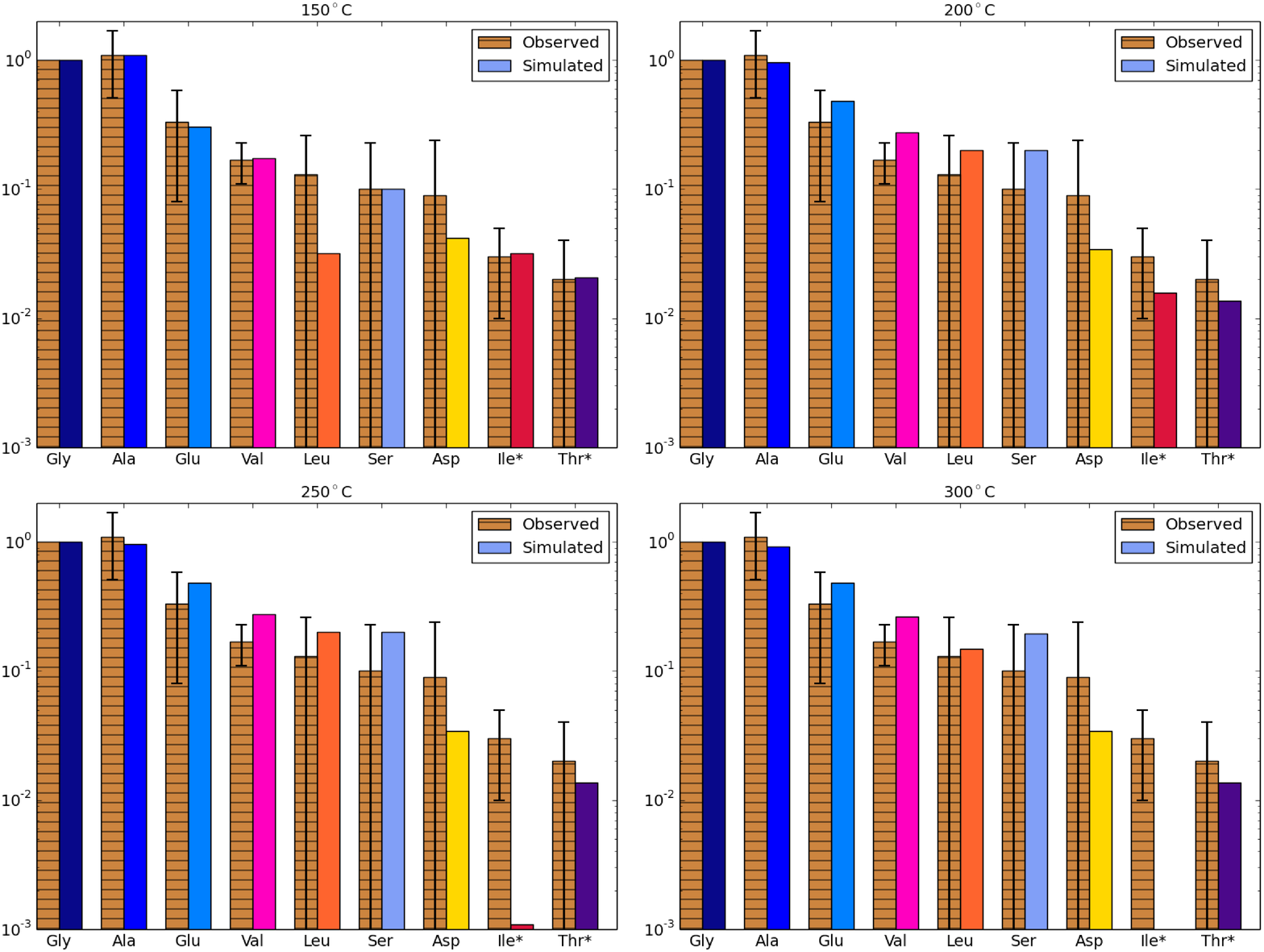}
\caption{Relative frequencies of amino acids in the CR2 subclass. Abundances of nine amino acids are shown relative to glycine, reported in moles of amino acid/moles of glycine. Observed frequencies are shown in brown with black horizontal stripes, and black error bars representing the standard deviation. The modeled frequencies are placed beside the observed frequencies with a unique solid color representing each amino acid. The four panels represent the progression of relative amino acid abundances with temperature, starting at $150^{\circ}$C and increasing by $50^{\circ}$C until $300^{\circ}$C. *Hydrolyzed amino acid abundances were not determined due to co-elutions, therefore the relative abundances represent those obtained from the unhydrolyzed samples (moles of free amino acid/moles of free glycine).}
\label{CRfreq}
\end{figure*}

The effects of this optimization are apparent in the excellent agreement between observed frequencies and the model frequencies at $150^{\circ}$C. All frequencies also remain well within the observed frequency error bars at $200^{\circ}$C, but as temperature increases further, we see a more dramatic change the frequency pattern. Threonine (thr) is the first amino acid to break the pattern, followed quickly by lysine (lys). Both amino acids produce concentrations less than 3*$10^{-3}$ that of glycine by $300^{\circ}$C. At this temperature, isoleucine (ile) is the only other amino acid to have a noticeable frequency degradation. All other amino acids retain modeled abundances akin to their observed frequencies throughout the temperature progression. This can be partially explained by thermodynamic favourability, as lys and ile have the highest Gibbs energies of formation of the 10 amino acids through temperature increase, and are therefore the least favourable (see Appendix I Figure~\ref{AppendixFigure1}).

\subsection{CR2 Meteorites}

Figure~\ref{CRfreq} shows the change in relative frequencies of nine amino acids with temperature among the CR2 meteorites. Relative frequencies are given in moles of amino acid/moles glycine. This time, each panel shows (in brown with black horizontal stripes) observed frequencies of amino acids relative to glycine among CR2 meteorites. Beside these observed frequencies stand (each in an individual color) the simulated relative frequencies of amino acids with respect to glycine. The amino acid lysine, shown in Figure~\ref{CMfreq}, is not shown for the CR2 equivalent, as there are no data among CR meteorites for lysine concentrations.

As for the CM2 subclass, we perform a chi-squared minimization on aldehyde abundances for the CR2 subclass. This is done by varying the initial aldehyde concentrations and measuring the chi-squared between the relative amino acid frequencies from simulation and the average relative amino acid frequencies in CR2 meteorites. The aldehyde abundances---in percent normalized to water---which minimize the chi-squared are shown in the CR column of Table~\ref{CMRconcs}.

These optimized aldehyde abundances produce model relative frequencies in good agreement with observed relative frequencies at $150^{\circ}$C (see Figure~\ref{CRfreq}). All model frequencies stand well within their observed frequency error bars. The pattern then changes rapidly with increasing temperature. Some relative abundances (glu, val, leu and ser) increase between $150^{\circ}$C and $200^{\circ}$C, while others (ala, ile and thr) decrease. Ile and thr have the largest decrease in relative abundance from $150^{\circ}$C to $200^{\circ}$C. Thr then stabilizes whereas ile decays to 10$^3$ by $250^{\circ}$C; and to only trace amounts by $300^{\circ}$C. Asp, and arguably ala, are the only amino acids to retain abundances similar to their observed frequencies throughout the shown temperature range.

\begin{figure*}
 \centering
 \includegraphics[width=0.8\textwidth]{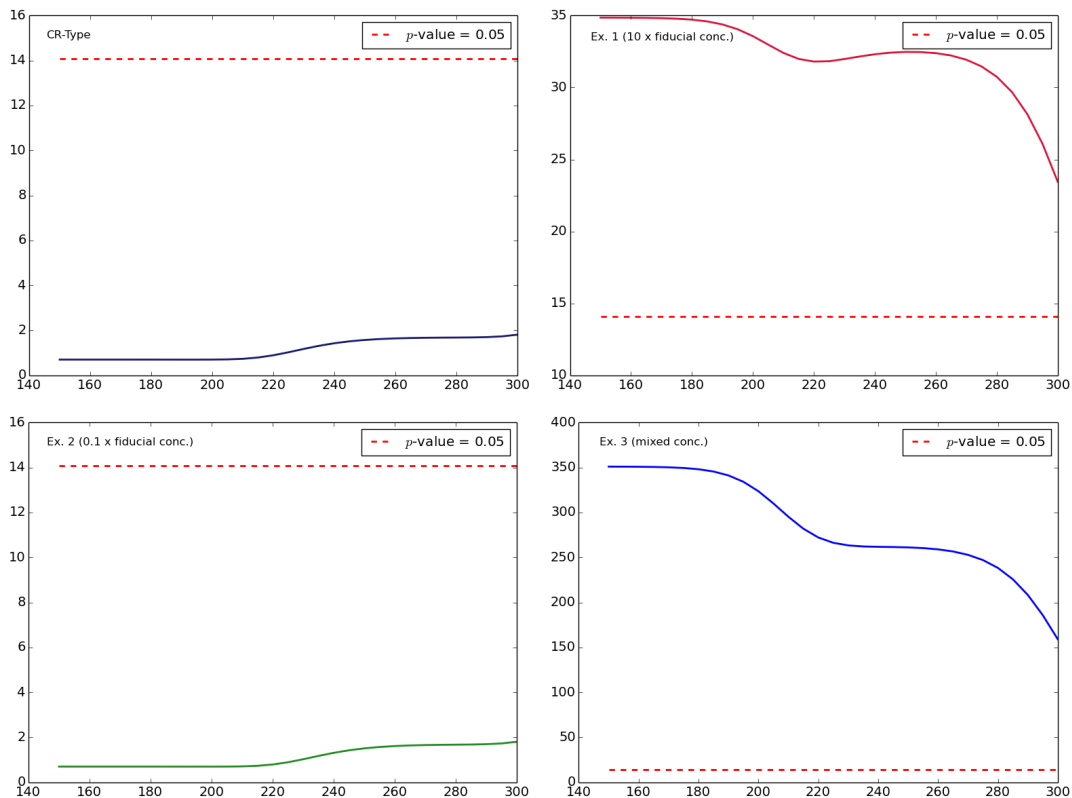}
 \caption{$\chi^2$ values for different ChemApp models, reflecting goodness-of-fit evolution for temperatures ranging from 150$^\circ$C to 300$^\circ$C, for four different initial aldehyde concentrations. The CR-Type plot represents the $\chi^2$ values for the aldehyde abundances shown in the CR column of Table \ref{CMRconcs} and used in Figure \ref{CRfreq}. The Ex. 1 plot represents the $\chi^2$ values for a model run with aldehyde concentrations an order of magnitude greater than those shown in the CR column of Table \ref{CMRconcs}. The Ex. 2 plot represents the $\chi^2$ values for a model run with aldehyde concentrations an order of magnitude lower than those shown in the CR column of Table \ref{CMRconcs}. The Ex. 3 plot represents the $\chi^2$ values for a model run with the formaldehyde, acetaldehyde and butryaldehyde concentrations shown in the CR column of Table \ref{CMRconcs}, and glycolaldehyde, propionaldehyde and pentanaldehyde concentrations greater by an order of magnitude. The red dotted line represents the $\chi^2$ value (from the $\chi^2$ distribution table) for a \emph{p}-value of 0.05 with 7 degrees of freedom. Any $\chi^2$ values lying above the dotted line represent a rejected null hypothesis.}
 \label{chichange}
 \end{figure*}

As these figures show frequencies relative to some baseline (in this case, glycine), it is entirely reasonable for future calculations to scale overall abundances of aldehydes. We assume some base concentration of formaldehyde, as this is the precursor for glycine, to which we normalize additional amino acids. It is the abundances of the aldehydes relative to formaldehyde that are responsible for the agreement between observed and modeled amino acid frequencies.

In the case of Figure~\ref{CRfreq}, the abundance of acetaldehyde is higher than its abundance in the CM2 model by approximately a factor of 2. This makes sense visually. Acetaldehyde is the aldehyde involved in the modeled synthesis of ala; and the CR frequency of ala is approximately twice as much as its CM counterpart.

\section{Discussion}

\subsection{$\chi^2$ Minimization}

The aldehyde concentrations listed in Table~\ref{CMRconcs} minimize $\chi^2$ over variations in both concentration and temperature. While this helps to constrain formation conditions of amino acids in planetesimals, we are left with no impression of which thermodynamic consideration has a larger impact on organic synthesis. There are three aspects in our thermodynamic model that may be manipulated: pressure, temperature, and composition. 

We set pressure to a typical value for the interior of parent bodies and hold it constant. The effect of pressure on Gibbs free energy before the liquid-to-gas phase transition is minimal in comparison to the effects of changing temperature, as discussed above.

With regard to which aspect, temperature or composition (which we treat, for now, as a simplified model of aldehyde abundances), has a more significant effect on relative frequencies, we consider the evolution of $\chi^2$ values. We consider the relative frequencies of amino acids in CR meteorites as a representative case. 

Figure~\ref{chichange} shows the evolution of $\chi^2$ with increasing temperature, for a variety of initial aldehyde concentrations. The separate plots represent different ChemApp models which were run with aldehyde concentrations listed in Table~\ref{CMRconcs}. The plot labeled `CR-Type' shows the $\chi^2$ values for the model we used in Figure~\ref{CRfreq}. Examples 1, 2, and 3 show $\chi^2$ evolution for three different combinations of aldehyde concentrations, all varying with respect to the CR-Type concentrations. The concentrations used in `Ex. 1' are one order of magnitude greater than those in the CR column of Table~\ref{CMRconcs}, and those used in `Ex. 2' are decreased by one order of magnitude. `Ex. 3' disrupts the relative abundance pattern; concentrations of three aldehydes (formaldehyde, acetaldehyde, butyraldehyde) are those in the CR column of Table~\ref{CMRconcs}, but the three least 
abundant aldehydes (glycolaldehyde, propionaldehyde, pentanaldehyde) have been increased by one order of magnitude. 

The red dotted line represents a $\chi^2$ value of 14.067. This value comes from the $\chi^2$ distribution table for a $\emph{p}$-value of 0.05, with n -- 1 = 7 degrees of freedom. (n represents the number of relative amino acid frequencies which the simulation can vary; see Figure~\ref{CRfreq}.) All $\chi^2$ values higher than this line represent a rejected null hypothesis, i.e., a statistical unlikelihood of obtaining those simulated relative abundances given the observed values.

The result of this simple $\chi^2$ comparison shows that initial aldehyde composition has a greater effect on amino acid frequencies than does temperature evolution. In the CR-Type and Ex. 2 cases, $\chi^2$ increases by approximately a factor of 3 over a $200^{\circ}$C change. In the case of Ex. 1 and Ex. 3, $\chi^2$ decreases by approximately a factor of 2 over a $200^{\circ}$C change. Compositional effects on the contrary are usually much larger. The change in $\chi^2$ between the CR-Type composition and Ex. 1 and 3 compositions at $150^{\circ}$C is nearing 2 and 3 orders of magnitude respectively. Only the Ex. 2 composition doesn't change in $\chi^2$ relative to the CR-Type composition. The similarity between the CR-Type and Ex. 2 curves illustrates that relative amino acid abundances to glycine tend not to change very much if you equally scale aldehyde concentrations down (as opposed to up) from the CR-Type concentrations.

Both the CR-Type and Ex. 2 curves also stay well below the null hypothesis rejection line---yielding statistically probable relative frequencies. The Ex. 1 and Ex. 3 curves on the other hand start and remain above this line. This further illustrates the sensitivity of increasing the initial aldehyde concentrations versus increasing temperature.

\subsection{Does the Onion Shell Model Account for Amino Acid Data?}

 \begin{figure*}
 \centering
 \includegraphics[width=\textwidth]{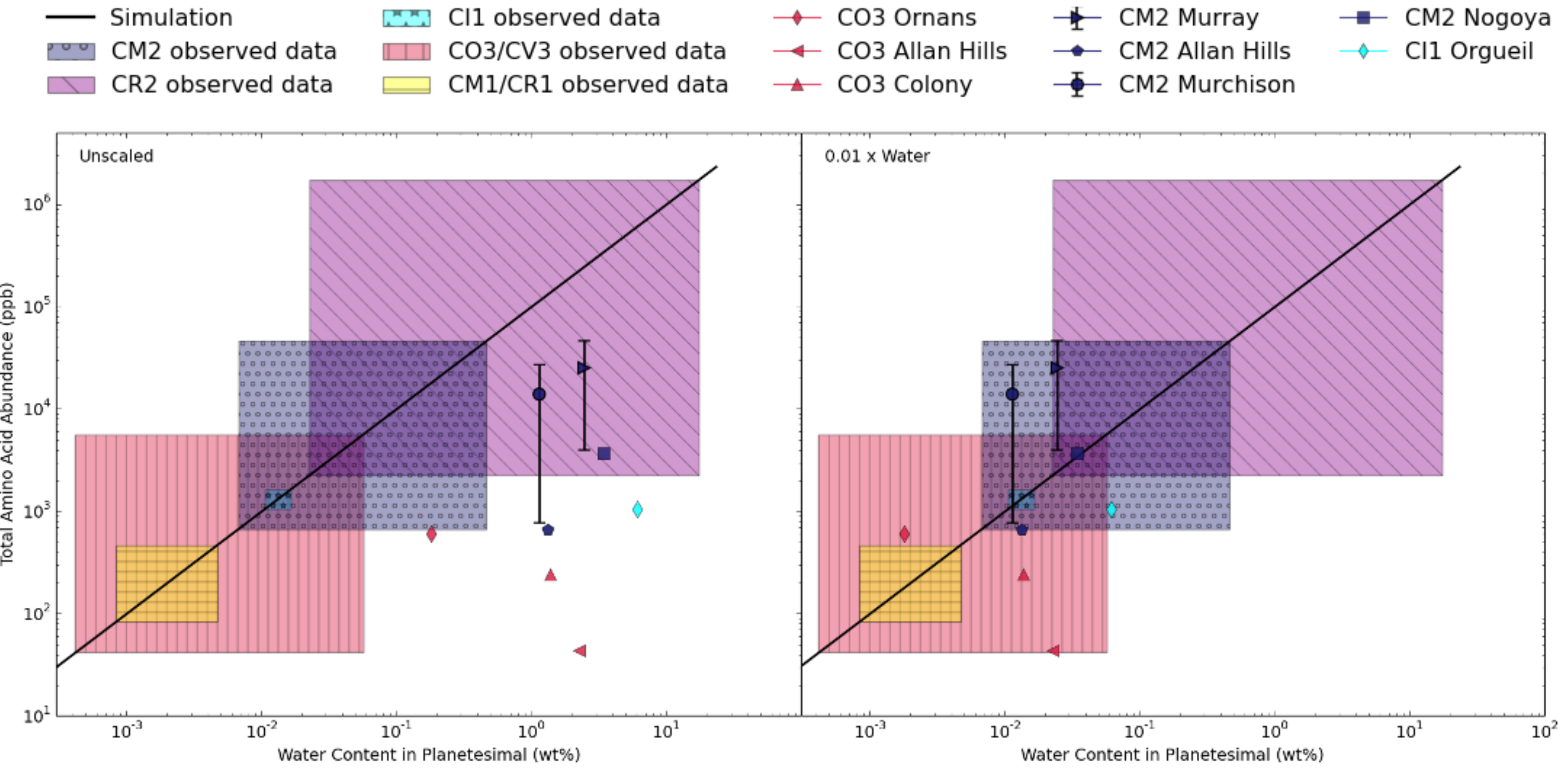}
 \caption{Theoretical total yield of amino acids in ppb with varying planetesimal water content. Water content varies in weight percentage over the entire planetesimal from 0\% to 50\%. Simulation is run with CR-like initial concentrations at 100bar and 100$^\circ$C. The shaded boxes indicate the observed range in amino acid abundance for each carbonaceous chondrite subclass (CI1, CM1/CR1, CM2, CR2, CV3/CO3). The individual data points represent the percent uncombined water content of separate samples of some of the carbonaceous chondrites used in this study, measured and/or reported by \citet{Jarosewich1990} and \citet{Wiik1956}. The percent uncombined water content of the data points in the left panel are as stated in the literature. The percent uncombined water content of the data points in the right panel are scaled down by a factor of 100.}
 \label{waterPercent}
 \end{figure*}

As discussed in Paper I, the onion shell model for planetesimals proposes that the carbonaceous chondrites within a certain subclass may share the same meteorite parent body \citep{Ehrenfreund2001, Glavin2011}, and that the different petrologic types within each subclass originate from different layers within the parent body; with each layer corresponding to a different temperature \citep{Weiss2013}.

In Figure~\ref{comparison}, the change in theoretical total amino acid abundances over the aqueous phase temperature range is too minute to support a simple temperature-differentiated onion shell model. For example the CR1 chondrites have total abundances just above $10^2$ ppb. If these meteorites are assumed to originate from locations within a parent body relatively closer to the core (whereas the CR2 meteorites, with total abundances near $10^5$ ppb, would be in the in-between layers associated with cooler temperatures), we would expect our CR2 models in Figure \ref{comparison} to move towards $10^2$ ppb as temperatures increase. Similarly, the CR3 chondrites, associated with the coolest alteration temperatures (pertaining to locations within the parent body nearest to the surface), also have a slightly lower abundance than the CR2 meteorites (near $10^4$ ppb for the only analyzed CR3 meteorite, QUE 99177). However, our CR2 model shown in Figure \ref{comparison} does not tend towards lower abundances at the highest temperatures (near 300$^{\circ}$C), or at the lowest temperatures (near 25$^{\circ}$C) thought to be in meteorite parent bodies of petrologic types 1--3 \citep{Morlok2013}. The lack of such behavior suggests that one of the main drivers of amino acid synthesis in planetesimals is the amount of water and organics in the planetesimal and not the temperatures inside each layer.

\subsection{Water Content and Amino Acid Abundance}\label{water}

Figure~\ref{waterPercent} explores the effect of varying water content further by showing the dependence of theoretical total amino acid yields to overall water percentage in the model planetesimal. The simulation is done with CR-like aldehyde concentrations, at 100 bar and 100$^\circ$C. These initial conditions are fairly arbitrary as Figure~\ref{comparison} shows how abundances are relatively the same across aqueous phase temperatures, and CR-like and CM-like initial compositions also produce similar curves.

The black curve in both panels of Figure~\ref{waterPercent} illustrates how total amino acid abundance can change by varying the water content of the fiducial model planetesimal.
The shaded boxes represent the minimum and maximum observed abundances in each of the carbonaceous chondrite subclasses, fit onto the simulation line. By drawing a line downwards from the bottom-left and top-right corners of each box, one could determine the theoretical minimum and maximum percent water content in the parent bodies hosting each of the carbonaceous chondrite subclasses.
By scaling the composition of a planetesimal from 0\% water and 100\% rock to 50\% water and 50\% rock, the simulated total amino acid abundances can be fit into the ranges of observed abundances for all the carbonaceous chondrite subclasses in this study. This result supports a composition-differentiated model; specifically, planetesimals with different overall water concentrations.

The aqua, dark blue, and red data points correspond to percent uncombined water measured and/or reported by \citet{Jarosewich1990} and \citet{Wiik1956} in separate samples (but the same body) of some of the carbonaceous chondrites that have been used in this study. Since the amino acid abundances and uncombined water content were not measured on the same sample, the amino acid abundance values for each data point correspond to the samples measured from the sources outlined in Paper I; with error bars corresponding to two or more sample measurements. The percent uncombined water values in \citet{Jarosewich1990} and \citet{Wiik1956} are measured by heating an uncovered sample to 110$^\circ$C for 1 hour and measuring the loss in weight. The measured values for these data points therefore consist of both excess free water from weathering and original free water. 

Notice in the left panel of Figure~\ref{waterPercent} how all of the data points lie to the right of the simulation line. This overabundance in measured uncombined water percent in comparison to our simulated values may be the result of substantial weathering. \citet{Bland2006} conclude that for a number of reasons, carbonaceous chondrites are more susceptible to weathering than other meteorite types and that even observed falls (as opposed to finds) may display the effects of significant weathering. The timescale for atmospheric, ground and/or rain water to be taken in by recently fallen meteorites could be very short: \citet{Pisani1864} dehydrated a sample of the Orgueil meteorite used in our study, and found that after only a few hours it had regained almost all the hygroscopic water that it had lost (9.15 wt\%).

In attempts to quantify the amount of weathering endured by each meteorite---and thus the corresponding excess free water in each sample---we look at the ratio of oxidized iron (Fe$_2$O$_3$) to metallic iron (Fe(m)). The oxidation of Fe(m) into Fe$_2$O$_3$ is a well known weathering effect \citep{Jarosewich1990}, therefore this ratio could provide a relative scaling factor for each of the water content values of our data points. Unfortunately this idea fails as none of our eight data samples yield non-zero and non-trace values for both oxidized and metallic iron.

In the right panel of Figure~\ref{waterPercent} we apply a constant scaling factor of 0.01 to all the uncombined water percent values. This allows us to estimate the ratio of original free water to excess free water from weathering that would allow the data points to coincide better with our simulation curve. Since individual samples will be subjected to distinct amounts of weathering, this 0.01 scaling factor is a simplification in attempts to typify the magnitude of excess water due to weathering in carbonaceous chondrites.
 
\subsection{Chemical Competition Between Amino Acids and Hydroxy Acids}\label{chemcomp}

So far we have, for simplicity,  focused on the formation of amino acids without discussing other competing chemical pathways.  This cannot be 
ignored in general, and we now turn to a discussion of one of the main competing processes.  
Hydroxy acids are similar to amino acids, except in place of the amino group (NH$_2$), they have a hydroxyl group (OH). The hydroxy acid analogous to glycine is glycolic acid, and the hydroxy acid analogous to alanine is lactic acid
Since hydroxy acids can also form from the Strecker synthesis pathway \citep{Schulte1995}, it is important to consider how the formation of hydroxy acids can reduce amino acid synthesis within meteorite parent bodies.  A key point about this chemistry is the role of ammonia that is so essential for Strecker synthesis.  
As NH$_3$ concentrations decrease, one expects that amino acid production will fall with respect to hydroxy acid synthesis \citet{Schulte1995}.

Table~\ref{hydroxy} lists the hydroxy acid to analogous amino acid meteorite abundance fractions for these molecules, from five meteorite samples. The molecular abundances are obtained from \citet{Cobb2014}, \citet{Pizzarello2010} and \citet{Monroe2011}. There is a large variation in the hydroxy acid to analogous amino acid fractions between the CI1, CM2 and CR2 meteorite subclasses. Within CR2 meteorites, there is only 0.04--0.09 times as many hydroxy acids as their analogous amino acids. In CM2 meteorites, hydroxy acids are 1--28 times more abundant than their analogous amino acids. And most significantly, in the Ivuna CI1 meteorite, hydroxy acids are 121--274 times as abundant as their analogous amino acids.

Within CI1 meteorite parent bodies, the molecular fractions suggest that conditions could be more favourable for hydroxy acid formation over amino acid formation. If hydroxy acid production is favoured within CI1 meteorite parent bodies, it could explain why our simulations produce almost 3 orders of magnitude more total amino acids than are present in CI1 meteorites. This suggests that observed NH$_3$ concentrations are positively correlated with amino acid synthesis within CI1 meteorite parent bodies, which is to be expected for Strecker reactions.  

Within CR2 meteorite parent bodies on the other hand, there appears to be only a small fraction of hydroxy acids relative to their analogous amino acids. This could mean that there is sufficient NH$_3$ within CR2 parent bodies allowing for optimal amino acid production. Any variation in amino acid abundances between the various CR2 meteorites could then be mainly from variations in overall parent body water content (as shown in Section~\ref{water}). This is validated by our simulations, which produce total amino acid abundances that agree well with the CR2 meteoritic abundances. 

\begin{deluxetable*}{ccccc}
\tablecolumns{5}
\tablecaption{Molecular fractions of hydroxy acids to their analogous amino acids and the ammonia content within one CI1 meteorite, two CM2 meteorites, and two CR2 meteorites. Data obtained from \citet{Cobb2014}, \citet{Pizzarello2010}, \citet{Monroe2011} and \citet{PizzarelloHolmes2009}.\label{hydroxy}}
\centering
 \tablehead{\colhead{Meteorite}&\colhead{Subclass}&\colhead{[mol glycolic acid]/[mol glycine]}&\colhead{[mol lactic acid]/[mol alanine]}&\colhead{NH$_3$ (nmol/g)}}
 \startdata
  Ivuna & CI1 & 120.88 & 274.17 & 5300\\
  Bells & CM2 & 16.06 & 28.00 & 280\\
  Murchison & CM2 & 0.68 & 1.94 & 1100\\
  GRA 95229 & CR2 & 0.09 & 0.05 & 18850\\
  LAP 02342 & CR2 & 0.04 & 0.04 & 14080\\[-3mm]
 \enddata
\end{deluxetable*}

The CM2 meteorites may represent the intermediate case between these two extremes. Here,  hydroxy acid to analogous amino acid fractions are in between the values for the CI1 and CR2 meteorites. NH$_3$ concentrations could therefore be moderately affecting amino acid synthesis within CM2 meteorite parent bodies, leading to similar abundances of hydroxy and amino acids.
This fits in well with the story thus far, as our simulations produce about an order of magnitude more amino acids than the CM2 meteoritic abundances, which could be explained by a moderate reduction of amino acid synthesis from relatively low NH$_3$ concentrations.

To investigate the possible role of reduced NH$_3$ as the ultimate cause of reducing amino acid synthesis within CI1 and CM2 meteorite parent bodies, we collected the available data on the ammonia content for each of the five meteorite samples in Table~\ref{hydroxy}. These abundances are obtained from \citet{Monroe2011} and \citet{PizzarelloHolmes2009}. Results show that the CR2 meteorites have over an order of magnitude more NH$_3$ than the CM2 meteorites, and about a factor of 3 more NH$_3$ than the CI1 meteorite. Since CR2 meteorite parent bodies have the greatest amount of NH$_3$, amino acid synthesis should be favoured over hydroxy acid synthesis \citep{Schulte1995}, which aligns well with our simulations and the meteoritic record. The CM2 meteorites however---especially in the case of the Bells sample---have less NH$_3$ than the CI1 meteorite. This does not fit perfectly with this picture, where amino acid abundances in CI1 meteorite parent bodies would seem to have been reduced more than in CM2 meteorite parent bodies, judging by the hydroxy acid to analogous amino acid ratios. This could perhaps partially be explained by the fact that water content within the petrographic type 1 meteorites tends to be higher than within the petrographic type 2 meteorites, which could have an effect on how diluted the ammonia is within the CI1 parent body.

\subsection{Amino Acid Decomposition}

One limitation of our model is that amino acid decomposition pathways are not simulated. Amino acids deaminate and decarboxylate in aqueous solutions at rates that vary with temperature \citep{McCollom2013,Sato,Qian}. High-pressure, high-temperature laboratory experiments infer through extrapolation that for temperatures less than $\sim$60$^{\circ}$C, some amino acids have half lives $>$ 10$^6$ years. These experiments also show that amino acids degrade rapidly at temperatures $\gtrsim$ 200$^{\circ}$C. However, since at lower temperatures, amino acid decomposition reaction rates may be longer than the aqueous lifetime of a meteorite parent body ($\sim$10$^6$ years) \citep{Travis2005}, it is reasonable to assume that not too much decay has occurred. The main exception would be if a meteorite parent body interior reaches temperatures $\gtrsim$ 200$^{\circ}$C, but as we have shown in Figures~\ref{CMfreq} and \ref{CRfreq}, simulated relative frequency patterns diverge from the meteoritic values at 250$^{\circ}$C. Hence, since we expect that interior temperatures never reach above 200$^{\circ}$C, decay effects should be minimal.

\subsection{A New Picture - Linking Amino Acid  Synthesis in Planetesimals to Disk Chemistry}

The results of our simulations suggest a new picture of amino acid synthesis within planetesimals. Total amino acid abundances within planetesimals remain relatively stable throughout the aqueous phase temperature range. This suggests that temperature and pressure are not the main driving factors for differences in amino acid abundances across meteorite types. We have already argued that hydroxy acid production could be favoured over amino acid production within certain meteorite parent bodies (e.g. CI1 and CM2), while amino acid production could be favoured over hydroxy acid production within others (e.g. CR2). This could be due to varying NH$_3$ to water levels within these parent bodies. Secondly, initial water content within the planetesimal, and the scaling organics within that water, could account for the variation of amino acid synthesis within meteorite parent bodies of the same subclass. This is demonstrated by the shaded boxes in Figure~\ref{waterPercent}, which show that varying the water content within CR2 meteorite parent bodies---whose amino acid abundances may not be curbed by low NH$_3$ concentrations---from 0.02--18 wt$\%$ (purple box), could account for the variation in CR2 meteoritic amino acid abundances. Amino acid synthesis within CI1, CM1, CR1, CO3 and CV3 meteorite parent bodies on the other hand, could first be reduced due to relatively low NH$_3$ concentrations, and then varied further by the ranges in their parent body water contents.  

The variation of water and ammonia within planetesimals is ultimately tied to where and how these bodies formed in their host protostellar disks. Astrochemistry models for such systems are starting to become sophisticated enough to allow us to explore this connection (see \citet{Pontoppidan}).

Our proposed picture conforms well with the water percentages measured in various chondrites in the literature. \citet{Weisberg2006} state that petrologic type 1 chondrites have the highest water content by weight (18--22$\%$), followed by type 2 chondrites (2--16$\%$) and then type 3 chondrites (0.3--3$\%$). Figure~\ref{waterPercent} shows that increasing the water content in the meteorite parent body leads to greater amino acid production. Since CI1 meteorite parent bodies appear to have had the largest water percentages, the reduction of amino acid production from low NH$_3$ concentrations within these parent bodies would have to be greater in order to produce the comparatively low amino acid abundances measured in CI1 meteorites. CR2 meteorite parent bodies on the other hand, which appear to have had mid-range water concentrations (2--16$\%$), have ranges in meteoritic amino acid abundances that can be explained almost completely with varying the water content within our simulations (from 0.2--18$\%$). CM1, CR1, CO3 and CV3 meteorite parent bodies would then require some level of amino acid abundance reduction (by having relatively low NH$_3$ concentrations), as well as an assortment of water contents, to account for their lower and varied meteoritic amino acid abundances.

In circumstellar disk models from \citet{Woitke2009} and \citet{Visser2012}, ice water content along the mid-plane is found to increase swiftly (but not instantaneously) at the ice line, after which it changes only minimally for several AU. The water content within each forming planetesimal varies as a function of its distance from the ice line. For the Solar System, the ice line during planet formation may have been somewhere within the asteroid belt, as modeled by \citet{Raymond2004} (at 2 and 2.5 AU from the Sun). Most of Earth's meteorites (including carbonaceous chondrites) are believed to originate from the asteroid belt. If the water ice distribution 
is inherited from the cicumstellar disk models explained above, then the proposed variation in water content between the parent bodies of the carbonaceous chondrites in our study could be possible.  

The variation of ammonia across protostellar disks has also been computed (see review  \citet{Pontoppidan}). It is known observationally that NH$_3$ can be
ruled out as a significant carrier of nitrogen in the warm inner regions of disks so that most of the nitrogen in the terrestrial planet forming region may be stored
as N$_2$.  At larger disk radii, nitrogen will be found in the form of ammonia, which explains its abundance in comets and in the atmospheres of giant planets.  Applied
to our model, this suggests that if high ammonia levels secure dominant amino acid abundances, then the parent bodies for such meteorites would have derived from 
such outer disk regions beyond the terrestrial planet formation region.

\section{Conclusions}
 
In general, we see good agreement between our thermodynamic computations and observed amino acid abundances. When applying our fiducial model, thermodynamics well explains the total amino acid concentrations we see in CR2 meteorites and the relative frequencies we see in both the CM2 and CR2 subclasses. 

Our models for CM-type and CR-type total amino acid abundances at temperatures less than $311.03^{\circ}$C remain near 7x$10^5$ parts-per-billion (ppb). This value is in agreement with the average observed abundance in CR2 meteorites of $4\pm7$x$10^5$, but approximately an order of magnitude higher than the average observed abundance in CM2 meteorites of $2\pm2$x$10^4$. These simulation results are not sensitive to slight changes in the initial aldehyde concentrations. For this reason we think the results will remain fairly similar regardless of whether relative amino acid abundances in hydrolyzed or unhydrolyzed meteorite extracts are used in calculations. An increase in the production of the competing hydroxy acids in the meteorite parent body can explain the discrepancy between the CM-type model computations and the observed amino acid abundances in CM2 meteorites.
 
There is a sharp drop in total amino acid abundance at the liquid-to-gas phase transition of the aqueous solution. We attribute this drop-off to the decline in thermodynamic favourability of synthesizing amino acids once in a gaseous phase mixture.

Our models do not decrease in total amino acid abundance by any significant amount over the course of the aqueous phase temperature range. This lack of declination does not support a simple temperature-differentiated onion shell model.

Instead, we propose a new explanation for amino acid abundances in planetesimals. In this new model, the production of hydroxy acids is favoured over the production of amino acids within certain meteorite parent bodies (e.g. CI1 and CM2). This is likely due to the low NH$_3$ concentrations within these parent bodies, which, as we have seen, results in lower amino acid production in preference to hydroxy acid production. (Protoplanetary disk astrochemistry models predict low ammonia levels in the inner regions of disks.) Then, to get the variation in the amino acid abundances within the same carbonaceous chondrite subclasses, the total water content (and scaling reactant content) could vary from parent body to parent body. 

Our model accounts for the relative frequency pattern in amino acids among CM2 and CR2 chondrites. The observed and theoretical frequencies with respect to
glycine match to well within an order of magnitude in both cases, at a temperature of $150^{\circ}$C. The degradation of lysine and threonine for CM-type relative frequencies and of isoleucine for CR-type relative frequencies by $250^{\circ}$C and $300^{\circ}$C, is in agreement with 3D models of interiors of young planetesimals; which are not heated much beyond $200^{\circ}$C. This change in amino acid frequencies may suggest a method to estimate internal meteorite parent body temperatures within that range.
 
These various aspects of the amino acid record in meteorites point towards the deep connection that likely exists with local disk astrochemistry. In particular, the processes that determine the abundances of ammonia, water, and aldehyde concentrations at different disk radii and  their incorporation into planetesimals ultimately provide the initial conditions  for Strecker synthesis of the amino acids. It is therefore possible to link biomolecule patterns in various types of meteorites to a much broader picture of disk astrochemistry.  
\newline

We would like to thank the anonymous referee, whose useful comments led to significant improvements in this paper. We are indebted to the excellent contributions of Jeff Emberson and Darren Fernandes who, during their undergraduate theses with REP at McMaster, helped 
establish some of the basic methods used in the early phases of this work.    
We would also thank Paul Ayers and his graduate student Farnaz Zadeh at McMaster University 
who helped immensely with their theoretical chemistry expertise. A.K.C. is very grateful to Mikhail Klassen and Alex Cridland for their assistance and prowess in Python and coding in general. A.K.C. and B.K.D.P. would like to thank the Canadian Astrobiology Training Program who supported them with CATP Graduate and Undergraduate Fellowships, funded by NSERC CRSNG. R.E.P. was supported by an NSERC Discovery Grant.

\section*{Appendix I - Additional Gibbs Energies of Formation}

\begin{figure*}
\centering
\includegraphics[width=\textwidth]{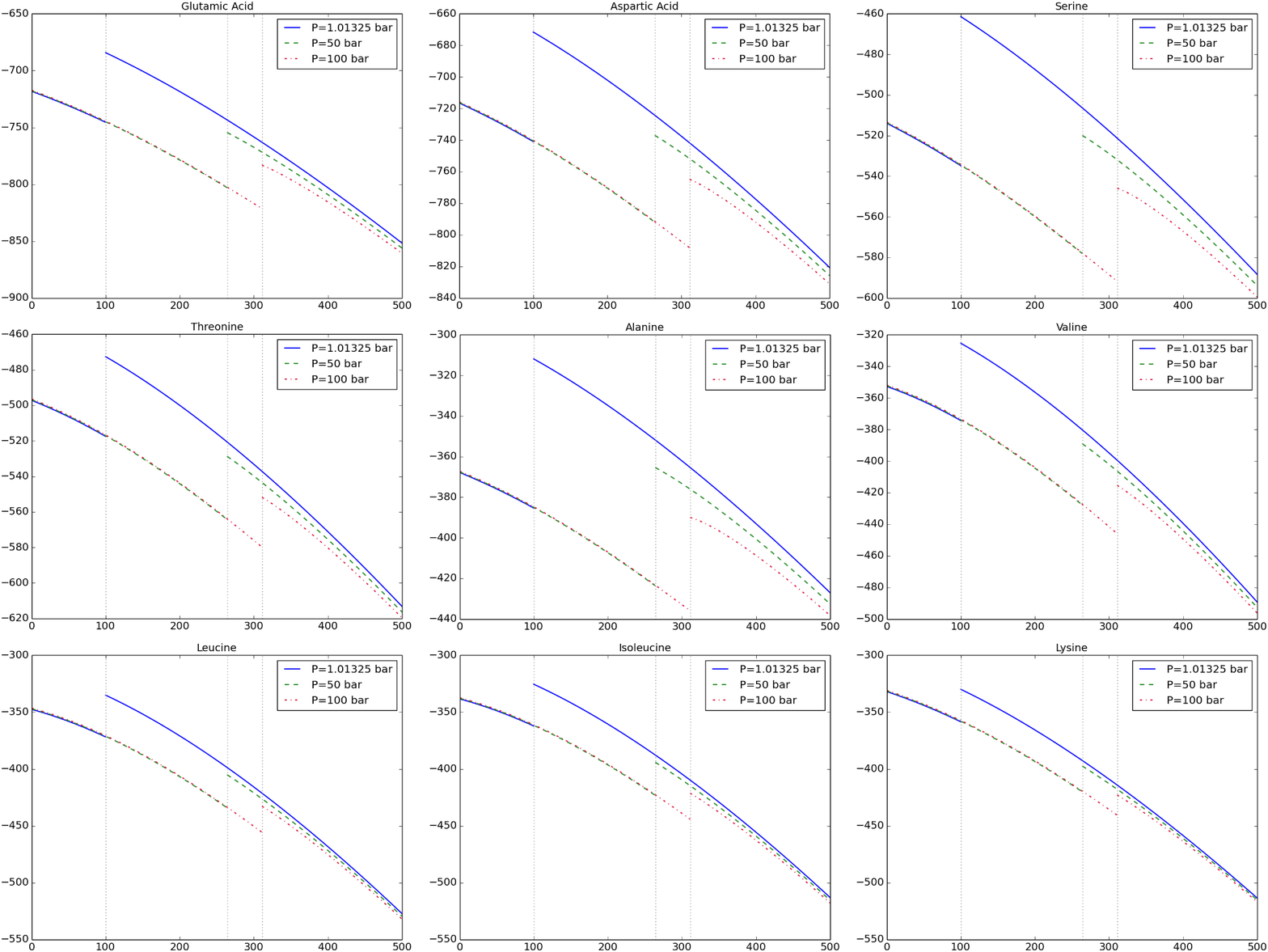}
\caption{Plots extending Figure \ref{pressure} for the nine other amino acids considered in this simulation, in order of increasing Gibbs free energy of formation (at 50$^{\circ}$C and 100 bar). This plot depicts each amino acid's change in Gibbs free energy of formation (kJ/mol) with respect to temperature ($^{\circ}$C), at various pressures. The vertical grey dotted lines represent the temperatures of the the aqueous-to-gas phase transitions of the amino acid (and hence of water) at 1.01325 bar, 50 bar and 100 bar---which are 100$^{\circ}$C, 263.97$^{\circ}$C and 311.03$^{\circ}$C respectively.}
\label{AppendixFigure1}
\end{figure*}

In Figure~\ref{AppendixFigure1} we extend Figure \ref{pressure} (Gibbs free energy of formation for Glycine) to the other nine amino acids considered in our simulation, and order them by increasing Gibbs free energy (at fiducial values of 50$^\circ$C and 100 bar). All amino acid curves are piecewise, and increase in Gibbs free energy at the point of discontinuity. Table~\ref{GibbsValues} displays exact values for the Gibbs free energies of formation for each of the proteinogenic amino acids used in our simulation, plus one non-proteinogenic amino acid, at the fiducial temperature and pressure.

It is worth noting that the order of amino acids in Table~\ref{GibbsValues} does not directly match the order of amino acids in Figure~\ref{CMfreq} or Figure~\ref{CRfreq}. This implies that the amount of each amino acid synthesized in planetesimals is dictated by more than just its energetic favourability. Additional factors, such as the concentration of each aldehyde precursor, cause a dispersion in the ordering of relative amino acid abundances.

\subsection*{$\alpha$-A\textlcsc{minobutryric} A\textlcsc{cid}}

\begin{figure}[ht!]
\centering
\includegraphics[width=80mm]{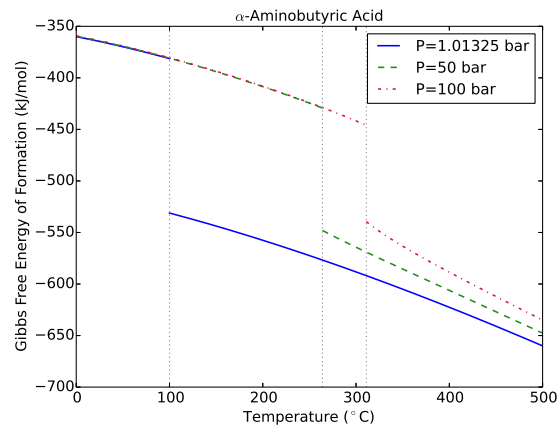}
\caption{A plot of Gibbs energy dependence on pressure and temperature for the non-proteinogenic amino acid $\alpha$-aminobutyric acid. Temperature is allowed to vary from 0$^{\circ}$C to 500$^{\circ}$C. Three curves are shown, representing the Gibbs energies at pressures of 1.01325 bar (blue curve), 50 bar (green curve), and 100 bar (red curve).}
\label{a-aminobutyric}
\end{figure}

Figure~\ref{a-aminobutyric} shows the Gibbs free energy of formation for $\alpha$-aminobutryric acid with respect to temperature, at three different pressures. In contrast to the plots in Figure~\ref{GibbsValues}, the Gibbs free energy of $\alpha$-aminobutryric acid decreases after the boiling points of water for each of the three pressure curves. This signifies an increase in energetic favourability after the liquid-to-gas phase transition. 

Since we have been basing our simulations on an aqueous phase chemical reaction (Strecker synthesis), comparing the Gibbs energy of $\alpha$-aminobutryric acid in the liquid phase to that of the proteinogenic amino acids is most important. This is shown in Table~\ref{GibbsValues}, where $\alpha$-aminobutryric acid ranks 7th in energetic favourability at 50$^\circ$C and 100 bar. From compiling amino acid abundances from the sources outlined in Paper I, $\alpha$-aminobutryric acid is found to be tied for 5th in highest average amino acid abundance in both the CM2 and CR2 subclasses. Observed $\alpha$-aminobutryric acid abundances therefore rank near what we would expect based on their relative Gibbs free energies. But as stated above, there are more factors to amino acid synthesis than just energetic favourability.

\begin{deluxetable}{p{35mm}p{35mm}}
\tablecaption{Gibbs free energies of formation for each of the proteinogenic amino acids used in our simulation plus one non-proteinogenic amino acid; at 50$^\circ$C and 100 bar. Amino acids are ranked by increasing Gibbs free energy of formation.\label{GibbsValues}}
\centering
 \tablehead{\colhead{Amino Acid}&\colhead{Gibbs Energy (kJ/mol)}}
 \startdata
Glutamic Acid & -730.0406\\
Aspartic Acid & -727.2654\\
Serine & -523.2076\\
Threonine & -505.6537\\
Glycine & -384.4235\\
Alanine & -375.3493\\
$\alpha$-Aminobutyric Acid\tablenotemark{*} & -369.0258\\
Valine & -361.3631\\
Leucine & -357.0445\\
Isoleucine & -347.7252\\
Lysine & -342.9651\\[-3mm]
 \enddata
 \tablenotetext{*}{Non-proteinogenic.}
\end{deluxetable}

\section*{Appendix II - Aldehyde Substitutions}

7 of the 10 amino acid synthesis simulations are performed using the Gibbs free energy data from aldehyde substitutes rather than the actual Strecker precursors listed in Table~\ref{aldehydes}. Due to the difference in steric configurations between the substitutes and the actual reactant aldehydes, we can assume that their Gibbs free energies may also differ. In order to try to identify the difference between their Gibbs free energies, we consulted the Ayers group in the Dept. of Chemistry and Chemical Biology at McMaster University, who performed an analysis using an electronic structure program called Gassian (F. Zadeh, P. Ayers and A. Patel 2015, personal communication).

From their analysis, it is discovered that the substitutes which have exactly the same atoms as the actual aldehydes, i.e., isobutryaldehyde (val), 3-methyl butanal (leu) and 2-methyl butanal (ile), have very similar Gibbs free energies to the actual aldehydes. Therefore the accuracy of the val, leu and ile synthesis simulations are likely not greatly affected by the aldehyde substitutions. The substitutes which do not have exactly the same atoms as the actual aldehydes, i.e., 3-oxopropanoic acid (asp), 5-aminopentanal (lys), 4-oxobutanoic acid (glu) and lactaldehyde (thr), were found to have a considerable difference in Gibbs free energy to the actual Strecker precursors.

In attempts to identify how greatly these four aldehyde substitutions might affect our amino acid synthesis results, we use the rough Gibbs data computed from Zadeh et al. (2015, personal communication) to adjust our aldehyde substitute Gibbs coefficients. We then rerun the CM-type simulations to compare to our previous results. We find that the reactions involving 3-oxopropanoic acid (asp) and 5-aminopentanal (lys) are not sensitive to the adjusted aldehyde Gibbs energies. 

Thus, there are only two reactions in this set, involving 4-oxobutanoic acid (glu) and 2-hydroxy-propanal (thr), that are quite sensitive to adjustments in Gibbs energies. We therefore conclude that the aldehyde substitutions used for the glutamic acid and threonine simulations may lead to significant deviations from accuracy, whereas the other five aldehyde substitutions are likely acceptable.

\newpage

\end{document}